\documentclass[twocolumn,prb,multicol,amsmath,amssymb,superscriptaddress]{revtex4}

\usepackage{graphicx}
\usepackage{dcolumn}
\usepackage{bm}
\usepackage{graphics}
\usepackage{epsfig,color}

\usepackage{physics}

\newcommand{\be}{\begin{equation}}
\newcommand{\ee}{\end{equation}}
\newcommand{\bea}{\begin{eqnarray}}
\newcommand{\eea}{\end{eqnarray}}

\usepackage[breaklinks,colorlinks,bookmarks=false,citecolor=red,linkcolor=blue,urlcolor=magenta]{hyperref}

\usepackage{xcolor}

\newcommand{\blue}[1]{#1}

\newcommand{\newrev}[1]{#1}

\begin{document}

\title{Canonically consistent quantum master equation for proton-transfer reactions}
\date{\today}
\author{Zahra Sartipi}
\email{zahra.sartipi@uni-potsdam.de}
\affiliation{University of Potsdam, Institute of Physics and Astronomy, Karl-Liebknecht-Str. 24-25, 14476 Potsdam, Germany\looseness=-1}

\author{Richard Gundermann}
\affiliation{University of Potsdam, 
Institute of Chemistry, 
Karl-Liebknecht-Str. 24-25,
14476 Potsdam, 
Germany}

\author{Janet Anders}
\affiliation{University of Potsdam, Institute of Physics and Astronomy, Karl-Liebknecht-Str. 24-25, 14476 Potsdam, Germany\looseness=-1}
\affiliation{Department of Physics and Astronomy, University of Exeter, Exeter EX4 4QL, United Kingdom}

\author{Peter Saalfrank}
\affiliation{University of Potsdam, 
Institute of Chemistry, 
Karl-Liebknecht-Str. 24-25, 
14476 Potsdam, 
Germany}

\begin{abstract}
The canonically consistent quantum master equation (CCQME) method to treat system-bath dynamics is used to describe intramolecular proton transfer in the thioacetylacetone molecule (TAA, C$_5$H$_8$OS), modeled as an $N$-level quantum system coupled to a solvent. The solvent is represented as a harmonic bath (a continuum of oscillators) characterized by an Ohmic-Drude spectral density. We benchmark \newrev{the secularized population dynamics and steady-state populations predicted by CCQME} against numerically exact hierarchical equations of motion (HEOM) theory and compare it to \newrev{the corresponding secularized Redfield results}.  Our results reveal that Redfield dynamics deviates increasingly from the HEOM reference as the system-bath coupling strength grows. \blue{In contrast, for not too strong couplings, the 
\newrev{secularized CCQME population dynamics} remains consistent with HEOM over an extended system-bath coupling range, and approaches the second-order mean force Gibbs state}. \newrev{A complementary non-secular calculation shows that retaining population–coherence coupling reveals limitations of the second-order treatment for coherence-sensitive observables.}
\end{abstract}

\maketitle

\section{Introduction}

Thioacetylacetone (TAA) is a typical $\beta$-thiooxoketone undergoing intramolecular proton transfer between its enol and
enethiol tautomers~\cite{Berg1983,Andresen2000,Posokhov2004}. This process is characterized by a
double-well potential energy surface (PES), where the lowest vibrational states are localized in their respective well and are
connected by quantum tunneling~\cite{Emsley1984,Doslic1999,Tatic2004}. In addition, the proton‑transfer dynamics are governed by coupling to an environment, i.e., a solvent. Within the condensed phase, especially with vibrational strong coupling~\cite{Fischer2023,Mandal2023}, the combined effects of proton tunneling, solvent changes, and light--matter interactions have attracted
significant interest as ways to control reaction rates and outcomes. To describe the proton-transfer dynamics in TAA accurately, an open-quantum-system approach formulated in terms of quantum master equations is required to account for the interactions between the molecule and its environment~\cite{Doslic1999}.

When the system-environment interaction is weak, open quantum systems are commonly described using second-order master equations, in particular the Redfield~\cite{Carmichael1999,Weiss2012,Redfield1965,BreuerPetruccione2007,CohenTannoudji1998} and the Lindblad--Gorini--Kossakowski--Sudarshan (LGKS) equations~\cite{Lindblad1976,GoriniKossakowskiSudarshan1976,BreuerPetruccione2007}. These approaches provide a computationally tractable framework for studying quantum dynamics in contact with thermal reservoirs~\cite{Becker2022,Lindblad1976,Redfield1965}, but both suffer from drawbacks. Specifically, the Redfield master equation fails to relax to the steady state that one expects a second-order master equation to achieve~\cite{Cresser2021,Becker2022}, particularly as the coupling strength increases~\cite{Thingna2012,Thingna2013}. Moreover, under weak to moderate coupling conditions, the Redfield master equation may produce unphysical negative populations in the density matrix~\cite{Becker2022,Hartmann2020,Fleming2011}. By contrast, 
the secular approximation of the Redfield master equation avoids such positivity violations. However, its equilibrium state is independent of the coupling strength, contradicting the principles of the Hamiltonian of mean force, which predicts population shifts arising from system--bath interactions~\cite{Becker2022,Jarzynski2017,Talkner2020}.
\par
To address these limitations, Becker \emph{et~al.}~\cite{Becker2022,BeckerDiss2022} developed an alternative master equation with a second-order canonical generator whose stationary state is the mean-force Gibbs state up to second order in the system-bath coupling~\cite{Trushechkin2022}, thereby providing a systematic correction of the traditional Redfield master equation. By construction, the CCQME achieves improved accuracy beyond weak-coupling approximations while requiring only the same level of computational complexity as Redfield calculations, making it valuable for applications spanning quantum optics~\cite{Carmichael1999,GardinerZoller2004}, chemical physics~\cite{Weiss2012,deVegaAlonso2017}, and quantum thermodynamics~\cite{Becker2022,BeckerDiss2022,Jarzynski2017,Talkner2020}.

In the present study, the CCQME formalism is applied to the TAA molecule interacting with a finite-temperature solvent, \blue{described by an Ohmic-Drude bath. The molecule is modelled by a one-dimensional asymmetric double-well potential. Using a vibrational-state representation of the system, we compute dynamical and steady-state populations over a wide range of system-bath coupling strengths~\cite{Trushechkin2022}, with Gibbs and second-order mean-force Gibbs states serving as equilibrium references. We start from the system ground and first excited states to model heating and cooling, respectively. 
 Furthermore, we present the dissipative dynamics of a Gaussian wavepacket initially moving towards the barrier to model a scattering-like event.} \newrev{By benchmarking the secularized population dynamics and stationary populations against} hierarchical equations of motion (HEOM) \cite{heom}, we specify the parameter regime where Redfield theory breaks down and show that the CCQME remains accurate \newrev{for these populations} up to {the onset of intermediate system-bath coupling~\cite{Cerisola_2024}. \blue{Our focus is on the steady-state, however, dynamics is also of interest.} 
\blue{TAA serves as a model system for a chemical reaction, in this case 
  a proton-transfer and underlying enol-enethiol tautomerism. This reaction and system have been well characterized computationally~\cite{Doslic1999,Tatic2004}, providing reliable reference points for addressing quantum-dynamical and open-quantum-system approaches, including master-equation descriptions. \blue{So far, thermodynamically consistent second-order generators such as that used in the CCQME have 
mostly been applied to simpler model systems, such as harmonic-oscillator-type models. The present work goes beyond this,}
 and thus} provides a controlled benchmark also for future studies of cavity-modified and solvent-dependent \blue{reactions, e.g., proton-transfer~\cite{Thomas2016,Fischer2023,Mandal2023,ExStateProtonTrans,Chen09_init1,Shi11_init2}}.

The paper is organized as follows: Sec.~\ref{Theory and formalism} provides the theory and formalism, Sec.~\ref{result} presents numerical results for population dynamics and steady states, as well as position expectation values. Sec.~\ref{conclusion} summarizes our conclusions and outlines perspectives for future work. Appendices~\ref{tabels}--\ref{WPDynBench} provide additional technical details.

\section{Model and Formalism}
\label{Theory and formalism}

\begin{figure}
    \centering
    \includegraphics[width=0.8\linewidth]{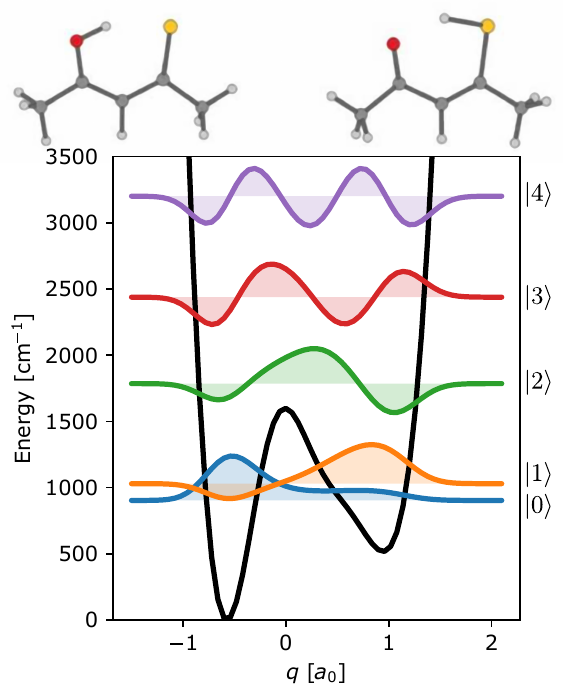}
    \caption{Enol (left) and enethiol (right) tautomers of TAA and corresponding PES along transfer coordinate $q$. The first five eigenstates are indicated. Note that the lowest state is localized in the left, and the first excited state is localized in the right well.}
    \label{FIG0}
\end{figure}

The system under consideration is the TAA molecule,  modeled by an effective one-dimensional transfer coordinate $q$ confined in a double-well potential $V_\text{TAA}(q)$, shown in Fig.~\ref{FIG0} with depicted tautomers corresponding to the minima. The molecule is linearly coupled to a solvent bath, modeled as a collection of independent harmonic oscillators. The total Hamiltonian is expressed as follows:
\begin{equation}
H_{\text{tot}} = H_\text{TAA}+ H_B + H_{SB}+H_{\text{ren}},
\label{tot_ham}
\end{equation}
where $H_{\text{TAA}}$ denotes the bare molecular Hamiltonian, $H_B$ the bath Hamiltonian, $H_{SB}$ the system–bath interaction, and $H_{\text{ren}}$ the renormalization term.

The bare TAA Hamiltonian representing the proton-transfer reads, using atomic units with $\hbar=1$, as 
\begin{equation}
H_{\text{TAA}} = -\frac{1}{2\mu_{\text{TAA}}}\frac{\partial^2}{\partial q^2} + V_{\text{TAA}}(q),
\label{eq:HTAA}
\end{equation}
where $\mu_{\text{TAA}}$ is the reduced mass and $V_{\text{TAA}}(q)$ is the double-well potential. Note that in Eq.~\eqref{eq:HTAA}, an original two-dimensional potential of Ref.~\citenum{Doslic1999} has been \blue{approximated by a one-dimensional form. This one-dimensional form had been 
 suggested in Ref.~\citenum{Fischer2023}}, where all relevant parameters referring to $H_\text{TAA}$ can be found. \blue{Moreover, Ref.~\citenum{Fischer2023} provides a justification 
  for the one-dimensional model, which closely resembles the lowest vibrational transition energy and the classical activation barriers of the original two-dimensional potential, this way leading to very similar system dynamics\cite{Fischer2023}.}
In the eigenbasis $\{|n\rangle\}_{n=0}^{N-1}$ of $H_\text{TAA}$, the proton-transfer coordinate $q$ can be written as
\begin{equation}
q = \sum_{n,m=0}^{N-1} q_{nm}\,|n\rangle\langle m|,\qquad
q_{nm}=q_{mn}^*,
\label{eq:q_operator}
\end{equation}
with matrix elements $q_{nm}=\bra{n}q\ket{m}$.
These matrix elements define the $N\times N$ matrix representation of $q$ in the truncated eigenbasis.
Table~\ref{tab:matrixelements} in Appendix~\ref{tabels} lists $q_{nm}$ for $N=6$ in atomic units, computed in the absence of system--bath coupling. 

The bath Hamiltonian in Eq.~\eqref{tot_ham} is modeled as a collection of independent harmonic oscillators ~\cite{BreuerPetruccione2007, HuPazZhang1992, Paz1994DecoherenceQBM, KarrleinGrabert1997, GrabertWeissTalkner1984} with mass $m_k$ and frequency $\omega_k$
\begin{equation}
H_B = \sum_k \left[\frac{p_k^2}{2m_k} + \frac{m_k\omega_k^2}{2}q_k^2\right],
\label{HB}
\end{equation}
where $q_k$ and $p_k$ are the position and momentum operators of the $k$-th bath mode, respectively.

The system–bath interaction 
takes the standard bilinear form~\cite{Doslic1999}
\begin{equation}
H_{SB} = q \otimes \mathbb{X},
\label{eq:HSB}
\end{equation}
with the collective bath coordinate expressed as
\begin{equation}
\mathbb{X} = - \sum_k c_k\, q_k,
\label{eq:X_operator}
\end{equation}
where $c_k$ are the coupling constants of the $k$-th bath mode.

The linear coupling to the bath induces a static modification of the  double-well potential $V_\text{TAA}(q)$, captured by the renormalization Hamiltonian
\begin{equation}
H_{\text{ren}} = \frac{1}{2}\,q^2\sum_k\frac{c_k^2}{m_k\omega_k^2}.
\end{equation}
This renormalization term is absorbed into the molecular Hamiltonian, defining an effective system Hamiltonian described as

\begin{equation}
H_S = H_{\text{TAA}} + H_{\text{ren}}.
\label{eq:HS}
\end{equation}
To obtain the effective system's energies, the Schrödinger equation
\begin{equation}
H_S \,\psi_n(q) = E_n\,\psi_n(q),
\label{eq:eigen_HS}
\end{equation}
 is solved numerically using a discrete variable representation (DVR) method \cite{Colbert1992}. More details are shown in the Appendix~\ref{NumericalDetails}. The eigenvalues for the lowest six energy levels are listed in Appendix~\ref{tabels}, Table~\ref{tab:eigenvalues},  and selected eigenfunctions are shown in Fig.~\ref{FIG0}, however without renormalization term. Apart from that, the renormalization is always included in this work. 

 TAA interacts with the solvent, and this interaction can be captured by a bath spectral density of Ohmic--Drude form~\cite{MayKuehn2011,BreuerPetruccione2007}

\begin{equation}
J(\omega)=\gamma\,\frac{\omega\,\omega_c^2}{\omega^2+\omega_c^2}, \qquad \omega>0,
\label{sspectral density}
\end{equation}
with coupling strength $\gamma$ and Drude cutoff frequency $\omega_c$ \blue{to be specified below.}

\par
In an Ohmic bath with a finite high-frequency cutoff ~\cite{HanggiIngoldTalkner2008, RitschelEisfeld2014} (Eq.~\eqref{sspectral density}), the decay timescale of the bath correlation function $t_{B}$ is much shorter than the intrinsic timescale of the system dynamics $t_S$ ($t_B\ll t_S$), so that bath fluctuations created by the system relax before the system noticeably changes, and therefore, they do not feed back onto the system at later times (Markovian regime). Under these conditions, the bath has only a short memory, and its influence on the system at time $t$ depends on the system state at the same time, captured by a time-local master equation. 
 \blue{Below we 
estimate that $t_B\ll t_S$ for the cases studied here. }
\par
Within the Born and Markov approximations~\cite{Redfield1965,BreuerPetruccione2007,Weiss2012,Carmichael1999}, the reduced density matrix $\rho(t)$ of the system is governed by the second-order Redfield master equation 

\begin{equation}
\frac{\partial{\rho}(t)}{\partial t}= -i[H_S,\rho(t)] + \mathcal{R}^{(2)}[\rho(t)]\,,
\label{redfield1}
\end{equation}
where the first term describes the unitary evolution governed by the effective
system Hamiltonian $H_S$ and the second term describes the dissipative contribution induced
by the environment, captured by the Redfield superoperator $\mathcal R^{(2)}$. 
In compact operator form, the latter can be written as
\begin{equation}
\mathcal{R}^{(2)}[\rho(t)]
=
\mathcal{K}\,\rho(t)\,q
- q\,\mathcal{K}\,\rho(t)
+ q\,\rho(t)\,\mathcal{K}^\dagger
- \rho(t)\,\mathcal{K}^\dagger q,
\label{eq:R2_def}
\end{equation}
where $q$ is the proton-transfer coordinate which enters the coupling operator given in Eq.~\eqref{eq:q_operator} and $\mathcal{K}$ is the
convolution operator (bath-memory integral) obtained in the asymptotic limit $t\gg t_B$, after replacing the upper integration limit in Eq.~\eqref{convoluted} by $\infty$. The bath correlation function and the tunneling rate $\mathbb{T}(\delta)$,  detailed in Appendix~\ref{TemporalCorrelationsBath} and
Appendix~\ref{TemperaturedependentTunneling}, respectively, enter the
evaluation of the convolution operator defined in
Eq.~\eqref{convoluted}. The derivation of the second-order Redfield master equation is given in Appendix~\ref{PPerturbative master equations},  Eq.~\eqref{eq:R2_def_time-dependent}.

We now turn to the CCQME approach as an alternative to the Redfield master equation. \blue{Here, we summarize
the formal CCQME framework developed in Refs.~\citenum{BeckerDiss2022, Becker2022}, providing a canonically consistent modification
of the Redfield master equation.}
\blue{In particular, the CCQME's dissipator is constructed such~\cite{BeckerDiss2022, Becker2022}, that the populations of its dynamical steady state match those of the so-called mean force Gibbs state, as we will review in the following.}

As a first step,  consider the total system, molecule and solvent,  relaxing to the global thermal (Gibbs) state~\cite{DAlessio2016,Deutsch2018}
\begin{equation}
\tau_{\mathrm{tot}}=\frac{e^{-\beta H_{\mathrm{tot}}}}{Z_{\mathrm{tot}}},
\qquad
Z_{\mathrm{tot}}=\tr_{SB}\!\left[e^{-\beta H_{\mathrm{tot}}}\right],
\label{eq:tau_tot}
\end{equation}
where $\beta = (k_{\mathrm B}T)^{-1}$ is the inverse temperature (with $T$ the solvent temperature and $k_{\mathrm B}$ Boltzmann's constant of the solvent), $Z_{\mathrm{tot}}$ is the partition function, and $H_{\mathrm{tot}}$ is given in Eq.~\eqref{tot_ham}. 
The corresponding equilibrium state of the molecule is obtained by tracing out the solvent degrees of freedom
\begin{equation}
\tau_{\mathrm{MF}}=\tr_{B}\!\left[\tau_{\mathrm{tot}}\right]. 
\label{eq:rhoMF_def}
\end{equation}
This reduced system state is the mean-force Gibbs state~\cite{Cerisola_2024,Cresser2021} (also called the reduced Gibbs state~\cite{BeckerDiss2022}).

The mean-force  Gibbs state expressed in
Eq.~\eqref{eq:rhoMF_def} can be perturbatively expanded in the system--bath interaction by using the
imaginary-time-ordered (Dyson) expansion of $e^{-\beta H_{tot}}$
\begin{equation}
\tau_{\mathrm{MF}}
=
\tau_{\mathrm{G}}+\Delta\tau_{\mathrm{MF}}^{(2)}+\mathcal O(H_{SB}^4),
\label{eq:rhoMF_series}
\end{equation}
where only even orders contribute. The first term
\begin{equation}
\tau_{\mathrm{G}}=\frac{e^{-\beta H_{S}}}{Z_S},
\qquad
Z_{S}=\tr_S\!\left[e^{-\beta H_{S}}\right].
\label{eq:rho0}
\end{equation}
is the Gibbs state (zeroth order)
in the limit of $H_{SB}\to 0$ (ultraweak coupling),
and the leading correction can be expressed through the second-order imaginary-time-ordered contribution $D^{(2)}$, expressed in Appendix~\ref{RedfieldSO_derivation} (Eq.~\eqref{eq:D2})~\cite{ BeckerDiss2022,Cresser2021}

\begin{equation}
\Delta\tau_{\mathrm{MF}}^{(2)}=\tau^{(2)}-\tau^{(2)}_{n},
\label{eq:rho2}
\end{equation}
where $\tau^{(2)}=\tau_{\mathrm{G}}\tr_B\!\big[\tau_B D^{(2)}\big]$ is the unnormalized second-order correction and $\tau^{(2)}_{n}=\tau_{\mathrm{G}}\tr_S\!\big(\tau^{(2)}\big)$ is the scalar normalization correction that guarantees
$\tr_S(\tau_{\mathrm{MF}})=1$ at order $H_{SB}^{(2)}$. Here, $\tau_B={e^{-\beta H_B}}/{\tr_B\!\left[e^{-\beta H_B}\right]}$ is the thermal equilibrium state of the bath.
 Thus, the second-order mean-force Gibbs state reads
\begin{equation}
\tau_{\mathrm{MF}}^{(2)}= \tau_{\mathrm{G}}+\Delta\tau_{\mathrm{MF}}^{(2)} ,
\label{mean-force-2}
\end{equation}
so that $\tau_{\mathrm{MF}}\simeq \tau_{\mathrm{MF}}^{(2)}$ up to $\mathcal O(H_{SB}^{(2)})$.

As shown by the equilibrium hierarchy in Appendix~\ref{RedfieldSO_derivation}, the second-order stationary condition (Eq.~\eqref{eq:stationarity_order2})
fixes the off-diagonal corrections, while in the
diagonal sector, it reduces to the population constraint and does not determine the
second-order population correction. The missing information enters only at the fourth order. 
Furthermore, reproducing mean-force Gibbs populations correctly up to $\mathcal O(H_{SB}^{(2)})$ requires, in general, fourth-order
generator information. 

The canonically consistent quantum master equation

\begin{equation}
\frac{{\partial}}{\partial t}\big[{\rho{(t)}}\big]_\text{CCQME}
=
-i[H_S,\rho(t)]
+
\mathcal R_t^{(2)}\!\Big[(1-\mathcal C_t^{(2)})[\rho(t)]\Big], 
\label{ccqme}
\end{equation}
provides an alternative way to achieve equilibrium populations relaxing to the mean-force Gibbs state up to $\mathcal O(H_{SB}^{(2)})$(Eq.~\eqref{mean-force-2}), without calculating $\mathcal R^{(4)}$. \blue{The derivation of Eq.~\eqref{ccqme} and
the details of the parameters are widely discussed in Refs.~\citenum{Becker2022, BeckerDiss2022}}. For further details see Appendix~\ref{PPerturbative master equations}. Eq.~\eqref{ccqme} is formulated in terms of the time-dependent Redfield superoperator $\mathcal R_t^{(2)}$ expressed in Eq.~\eqref{eq:R2_def_time-dependent}, i.e.\ the Redfield superoperator before taking the asymptotic Markovian limit used in Eq.~\eqref{redfield1}. The map $\mathcal C_t^{(2)}$, defined in Eq.~\eqref{eq:Ct2_def_integrated}, is a second-order canonical correction acting on system states. It supplies the missing stationary information in the diagonal sector through the contraction $\mathcal R_t^{(2)}\mathcal C_t^{(2)}$, so that in the asymptotic limit, where $\mathcal R_t^{(2)}\to\mathcal R^{(2)}$ and $\mathcal C_t^{(2)}\to\mathcal C^{(2)}$, Eq.~\eqref{ccqme} relaxes to the second-order mean-force Gibbs state (Eq.~\eqref{mean-force-2}), with $\mathcal C^{(2)}[\tau_{\mathrm G}]=\tau^{(2)}$, while the normalization term is included in Eq.~\eqref{eq:rho2}.
\par
 The time-independent superoperator ${{\mathcal{C}}^{(2)}}$ can hence be computed from the second-order mean-force Gibbs state,
 obtained from the tunneling rate $\mathbb{T} (\delta_{nm})$ (Eq.~\eqref{tunnelingrate_matsubara}) and its derivative (Eq.~\eqref{derivetive_matsubara}). Introducing projectors onto the off-diagonal ($\Pi_{\text{coh}}$) and diagonal ($\Pi_{\text{pop}}$) subspaces, 
a canonically consistent choice for the second-order map can be written as~\cite{Becker2022,BeckerDiss2022}
\begin{align}
\begin{aligned}
\mathcal C^{(2)}[\rho]
&=
\Pi_{\text{coh}}\frac{\mathcal R^{(2)}[\rho]}{i\delta}\\
&+ \Pi_{\text{pop}}
\sum_{n,l}
|q_{nl}|^2
\Bigg[\frac{\partial}{\partial \delta_{nl}}\,\Im\!\big[\mathbb{T}(\delta_{nl})\big]\
\mathcal D(L)[\rho]\\
&+ \Im\!\big[\mathbb{T}(\delta_{ln})\big]\;
\ket{n}\!\left(\frac{\partial \rho_{nn}}{\partial E_{n}}\right)\!\bra{n}\Bigg].
\label{ct}  
\end{aligned} 
\end{align}
 \blue{As mentioned, the derivation of the above second-order canonical correction map
follows 
Refs.~\citenum{Becker2022,BeckerDiss2022}. In the present work, we adopt this
formal framework and specialize it to the TAA proton-transfer model by using
the system Hamiltonian, coupling operator, and Ohmic-Drude bath parameters
specified below. All components entering Eq.~\eqref{ct}, including the
definitions of $\Pi_{\text{coh}}$, $\Pi_{\text{pop}}$, $\delta_{nl}$, the
derivative terms, and the dissipator $\mathcal D(L)$, are defined in
Appendix~\ref{Redfieldfulll_derivation}, where we present an illustrative
$N=2$ truncation of TAA and derive $\mathcal R^{(2)}$,
$\mathcal C^{(2)}$, and the corresponding CCQME.}
 \par
 
The theoretical development of the CCQME in Sec.~\ref{Theory and formalism} and the appendices is formulated in terms of the full non-secularized Redfield superoperator. However, for numerical results presented in Secs.~\ref{sec3a}-\ref{sec3c},  we \ make the widely used secular approximation~\cite{MayKuehn2011,Nitzan2006} for both Redfield and CCQME. \blue{This is because here we are predominantly interested in the populations and both master equations produce population artifacts in the non-secular case.   Redfield can give negative populations, while the modified dissipator of the CCQME introduces higher order overcorrections to the steady state coherences, which feed back into the dynamics of the populations via coherence-population coupling. These effects are suppressed within the secular approximation applied below.} 
\blue{However, we also perform a representative non-secular calculation and discuss limitations of the adopted methods separately in Sec.~\ref{limit}.}

\section{Results and Discussion}
\label{result}

\begin{figure*}
    \centering
    \includegraphics[width=1\linewidth]{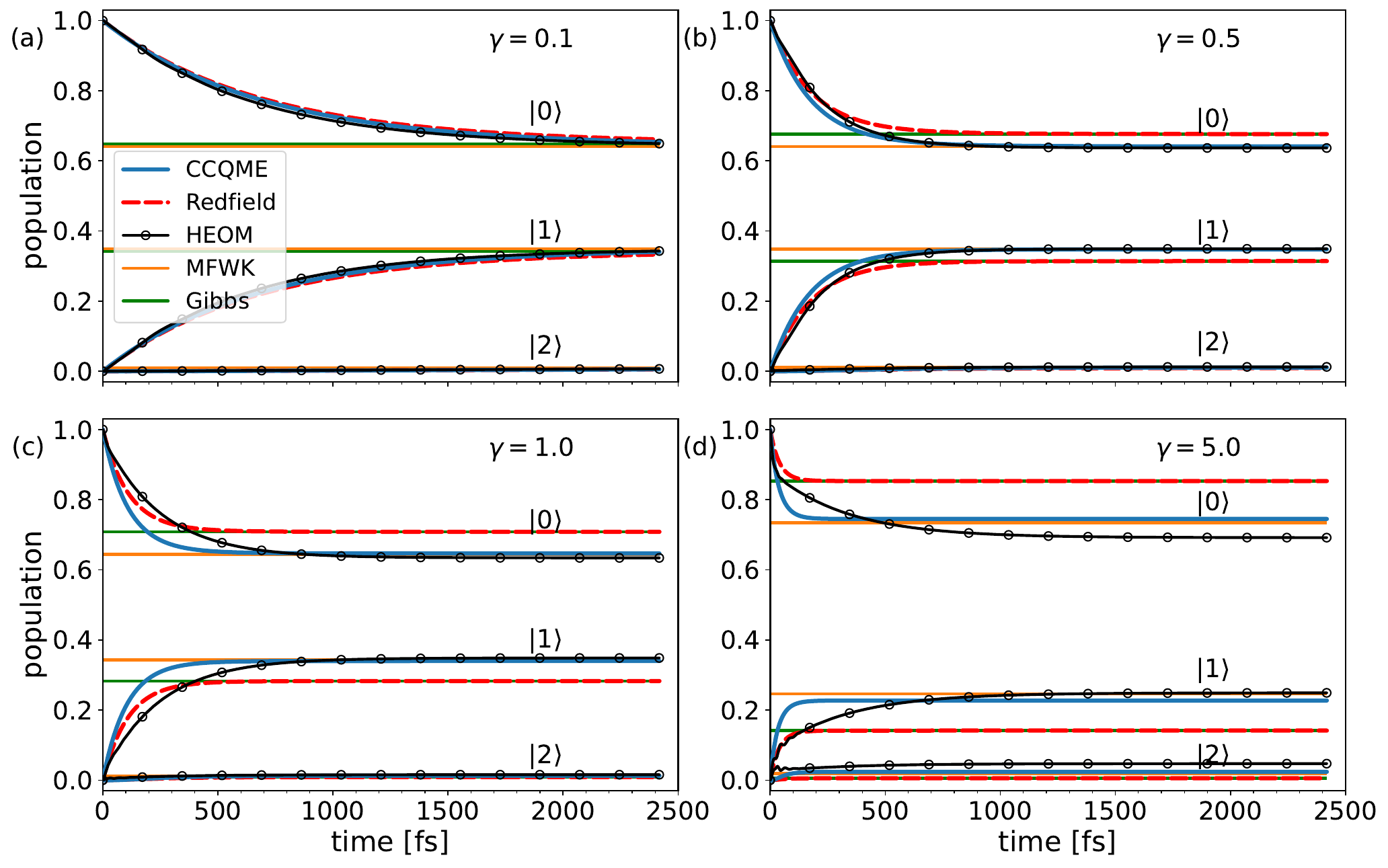}
\caption{Population dynamics of the three lowest eigenstates of the TAA proton-transfer coordinate at $T=300~\mathrm{K}$ for different system--bath couplings $\gamma$ up to $2.5~\mathrm{ps}$ computed with HEOM (black symbols), CCQME (blue solid), and Redfield (red dashed). Horizontal lines indicate the Gibbs (green) and second-order mean-force Gibbs (orange) stationary populations. The system was initially in the ground state. \newrev{Here, both CCQME and Redfield are computed within the secular approximation.}}
    \label{FIG1}
\end{figure*}

\begin{figure}
    \centering
    \includegraphics[width=1.0\linewidth]{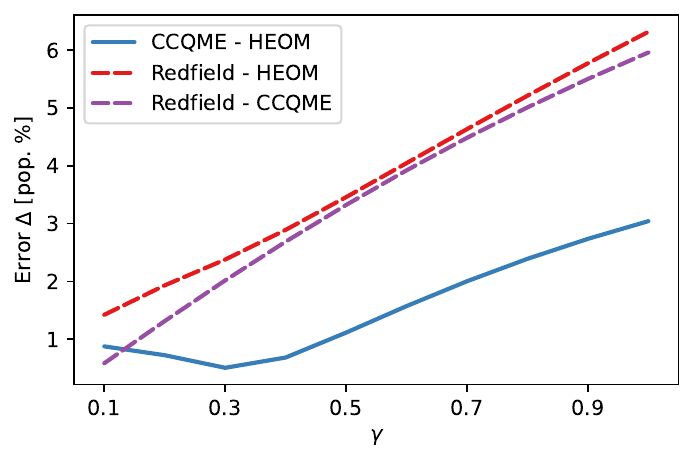}
    \caption{Time-averaged ground-state population error $\Delta$ (in \%) as a function of the system-bath coupling strength $\gamma$, \blue{in a range between 0.1 and 1.0, } comparing the CCQME and Redfield predictions with the numerically exact HEOM benchmark. The blue solid curve shows $\Delta$ (CCQME--HEOM), the red dashed curve shows $\Delta$ (Redfield--HEOM), and the purple dash--dotted curve shows $\Delta$ (Redfield--CCQME). The errors are averaged over the propagation time window $0$--$2.5$ ps corresponding to the dynamics in Fig.~\ref{FIG1}.}
    \label{fig:pop_error}
\end{figure}

\begin{figure*}
    \centering
    \includegraphics[width=1\linewidth]{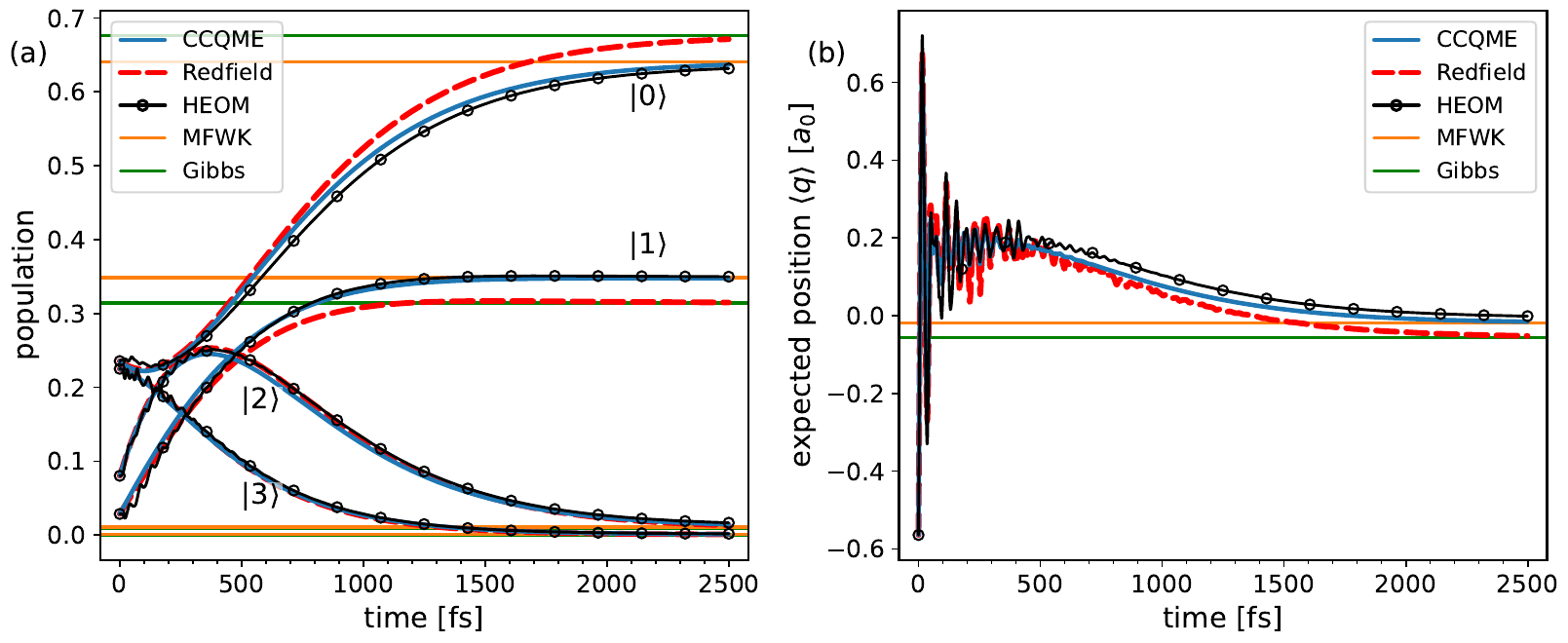}
    \caption{(a) Population dynamics of the four lowest eigenstates and expectation value of transfer coordinate $q$ for \blue{the Gaussian wavepacket Eq.~\eqref{eq:init_wavepacket}} as initial state, with coupling strength $\gamma=0.5$. (b) Corresponding coordinate expectation values $\expval{q}(t)$. \newrev{Both CCQME and Redfield are evaluated within the secular approximation.} \blue{Horizontal lines indicate the Gibbs (green) and second-order mean-force Gibbs (orange) stationary values.}}
    \label{fig:wp_dyn_g05}
\end{figure*}

This section examines the application of the \newrev{secularized} CCQME to the TAA proton-transfer system, in the presence of an Ohmic–Drude solvent bath, and compares it with the \newrev{secularized} Redfield master equation. As a benchmark, we employ HEOM, providing a numerically exact reference solution for this model. 
Details of the HEOM implementation are provided in Appendix~\ref{NumericalDetails}. \blue{\newrev{ In Secs.~\ref{sec3a}-\ref{sec3c}, we apply the secular approximation, while in Sec.~\ref{sec3d}, we \newrev{explicitly} consider the non-secular case.}} 

The analysis focuses \blue{mainly} on how these approaches capture both the transient population dynamics and the steady-state populations as the coupling strength $\gamma$ is varied. 
In what follows, we  consider a bath with a temperature $T=300 \ \text{K}$ \blue{and a cutoff frequency $\omega_c = 1.686 \cdot 10^{-3} \ E_h/\hbar$, corresponding to ca. 370 cm$^{-1}$. For comparison, the lowest Bohr frequency of the system, $\omega_{10}=E_1-E_0=127$ cm$^{-1}$, cf. Tab.~\ref{tab:eigenvalues} in Appendix A. An estimated bath correlation time as mentioned above is $t_B=1/\omega_c \sim $ 14 fs, compared to a characteristic system 
 time $t_S = 2\pi/\omega_{10} \sim$ 264 fs.} \newrev{The estimate $t_B \ll t_S$  indicates that the bath memory decays much faster than the characteristic motion of the TAA system, supporting the Markov approximation and the use of a time-local master equation.}
 We use three different initial states: (a) A situation where, initially, the system is in state $\ket{0}$ (ground state), localized in the left well, i.e., the system will be heated by interaction with the bath,  (b) an initial state  $\ket{1}$ (first excited state), localized to the right, where the system is effectively cooled down; and (c) a Gaussian wavepacket, initially moving from left to right.

\subsection{Ground state initialization} 
\label{sec3a}
The TAA system is initially in the ground state, $\rho(0) = |0\rangle\langle 0|$ and \blue{evolves}  toward a stationary state with \blue{equilibration} times ranging from several hundred femtoseconds to one picosecond.  Fig.~\ref{FIG1} illustrates the population dynamics of the three lowest eigenstates
for the TAA proton–transfer system, up to \(2.5\,\text{ps}\), for different values of the coupling strength \(\gamma\) \blue{between 0.1 and 5 $\ m_eE_h/\hbar$ (atomic units)}, using $N = 6$ energy levels. \blue{The populations are the diagonal elements of the reduced (system) density matrix.}  The evolution is computed with HEOM (black symbol lines), the CCQME (solid blue lines), and the Redfield equation (red dashed lines). Within the secular approximation with either Redfield or CCQME, the off-diagonal elements of the reduced density matrix, \blue{i.e., the coherences},  remain strictly zero for this particular initial state. The horizontal lines indicate
the stationary populations predicted by the Gibbs state (green line) given in Eq.~\eqref{eq:rho0} and by the second-order mean–force Gibbs state (orange line) given in Eq. \eqref{mean-force-2}.

From Fig.~\ref{FIG1}(a), it is evident that for $\gamma = 0.1$ all three \blue{dynamical} approaches show almost identical behavior. (Units for $\gamma$ are not explicitly indicated anymore from now on.) \blue{To put the coupling strength into context, 
 from Fermi’s Golden Rule we note that the ratio of the natural linewidth $\Delta E$ (for the lowest-energy transition) to the transition energy $\omega_{10}$ for a given coupling strength is estimated as $\Delta E/  \omega_{10} = 2 \abs{q_{10}}^2 \gamma$. For a coupling strength of $\gamma = 0.1$, this ratio is roughly 0.02, and 0.2 for $\gamma=1.0$, for example. For $\gamma=0.1$, which we will classify later as ``weak coupling'', the} populations evolve monotonically \blue{within about one ps} from the initially localized ground state toward the steady state, where \(\ket{0}\) and \(\ket{1}\) carry almost the entire probability, while the population of \(\ket{2}\) remains at the few-percent level and higher states are negligible. 
 HEOM, CCQME, and Redfield all converge to approximately the same steady state, reflecting that for this regime the traditional Redfield master equation is both dynamically accurate and thermodynamically consistent for the TAA subsystem.

For $\gamma = 0.5$ (Fig.~\ref{FIG1}(b)), \blue{
the populations equilibrate more rapidly, on a sub-ps timescale}. \blue{Clear differences between the Redfield dynamics and the HEOM/CCQME results become visible. Redfield populations converge to the Gibbs state asymptotically (Eq.~\eqref{eq:rho0}), thereby overestimating the ground-state population and underestimating the first excited-state population in comparison to the second-order mean-force Gibbs state. In contrast, CCQME  stationary populations coincide with the second-order mean-force Gibbs state (Eq.~\eqref{mean-force-2}), deviating from the Gibbs state.}

\blue{
For $\gamma = 1.0$ (Fig.~\ref{FIG1}(c)), the differences between Redfield and {HEOM} results are even more pronounced, especially during the intermediate-time equilibration dynamics, {while the CCQME performs significantly better
against HEOM.} Redfield converges to the Gibbs state as before. The HEOM dynamics equilibrates close towards the second-order mean-force Gibbs state, and the CCQME dynamics  settles to the second-order mean-force Gibbs state for both the population of \(\ket{0}\) and the population of \(\ket{1}\), as expected from the canonical correction. Thus, at this coupling strength, the CCQME still captures the asymptotic populations significantly better than Redfield. Note that CCQME is made to correct the steady-state populations up to second order in system-bath coupling, not necessarily the dynamics itself, which is clearly seen in the figure.
}

\blue{For the strongest coupling considered, $\gamma = 5.0$ (Fig.~\ref{FIG1}(d)), the deviations among the three approaches become largest, both in the transient dynamics and in the stationary state. Redfield theory remains close to the Gibbs state.
The HEOM reference, instead, remains much closer to the second-order mean-force Gibbs populations. The CCQME continues to follow the HEOM relaxation much more closely than Redfield, although a small but visible deviation between the CCQME stationary populations and the second-order mean-force Gibbs state starts to appear.

This discrepancy indicates that $\gamma = 5.0$ lies beyond the perturbative validity regime of the second-order CCQME, where higher-order finite-coupling corrections can no longer be neglected. \newrev{For this largest coupling, coherent oscillation and relaxation timescales are no longer as clearly separated. We therefore use the $(\gamma=5.0)$ result primarily as a demanding test case for the second-order secularized treatment, rather than as a parameter regime where all approximations are expected to be equally well controlled.} For further details, see Appendix~\ref{PPerturbative master equations}. 

\blue{To further classify the chosen coupling strengths, we characterize them} in the same manner as in Ref.~\citenum{Cerisola_2024}, \blue{using relative ground-state population errors}. 
Fig.~\ref{fig:pop_error} illustrates these population errors as a function of the coupling strength $\gamma$, ranging from 0.1 to unity, for Redfield, CCQME, and HEOM dynamics shown in Fig.~\ref{FIG1}. The errors are computed as time averages up to \(2.5\)~ps. 
If the relative error between the regular and second-order mean-force Gibbs state is below $\varepsilon =4 \cdot 10^{-3}$, this regime is identified as ``ultraweak coupling''(UW). Above this value, it is identified as ``weak coupling'' (WK). 
Furthermore, if the second-order mean-force Gibbs state deviates more than $\varepsilon$ from the steady state obtained from HEOM, \blue{the regime is characterized as ``intermediate coupling'' (IM).} Using those definitions, we find $\gamma=0.1$ is already in the WK regime with a relative error of $1.2 \cdot 10^{-2}$ and the IM regime, starts at $\gamma=0.4$. The latter also corresponds with the above mentioned deviations of CCQME from HEOM, after passing $\gamma = 0.5$. \blue{(We also performed another calculation with $\gamma=0.01$ where we enter the UW regime with an error of $1.2 \cdot 10^{-3}$ between the regular and mean-force Gibbs state (not shown).)}

From the red dashed curve in Fig.~\ref{fig:pop_error}, one can observe that the time–averaged error between the Redfield and the exact HEOM approach grows almost linearly as \(\gamma\) varies, 
reaching around six percent at \(\gamma=1.0\). This reflects that the traditional Redfield master equation is justifiable only in the weak–coupling limit and becomes less reliable as the system–bath interaction strengthens. The blue curve compares CCQME with HEOM. For couplings, \(\gamma \lesssim 0.6\), the CCQME error stays below 1.5 percent and even reaches a minimum around \(\gamma \approx 0.4\text{--}0.5\), where the CCQME and HEOM ground–state populations are almost identical. For couplings, \(\gamma \gtrsim 0.6\), the CCQME error begins to grow significantly. This means that higher–order corrections would be required in this regime. Nevertheless, even in this coupling regime, the
CCQME shows results closer to HEOM than those obtained
with the Redfield equation. The purple dash-dotted curve \blue{compares Redfield and CCQME and} grows strongly with $\gamma$, indicating that
the two master-equation predictions increasingly diverge as coupling strength increases,
mainly due to their different steady-state targets (regular vs. mean-force Gibbs). \blue{In passing we note that a classification of coupling regimes based on population errors is somewhat arbitrary, and other criteria such as linewidths (see above) or relaxation times (see below) can be helpful. }

\begin{figure*}
     
        \centering
        \includegraphics[width=1\linewidth]{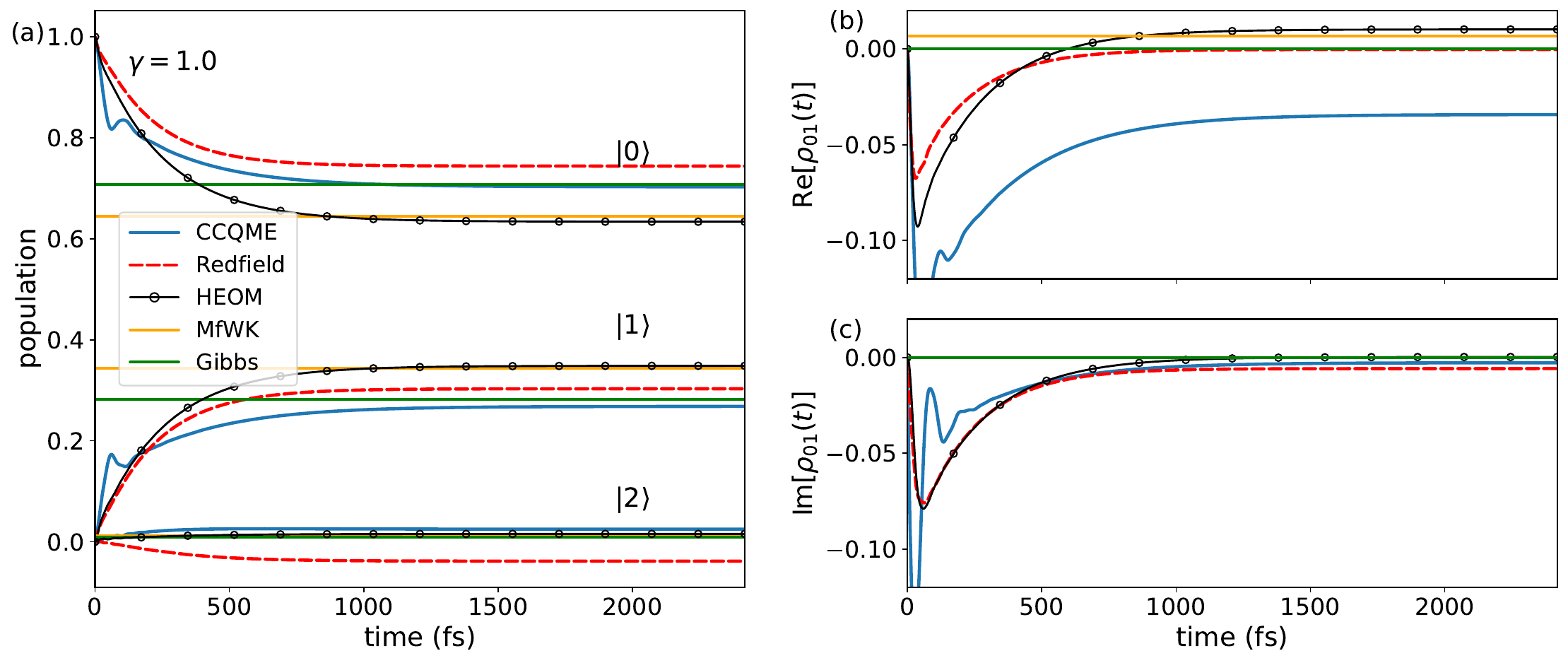}
        \caption{\blue{Non-secular population and coherence dynamics of the TAA subsystem for $\gamma=1.0$, using the ground-state initial condition. Panel (a) shows the populations of the three lowest energy eigenstates obtained from CCQME, Redfield, and HEOM, together with the standard Gibbs and second-order mean-force Gibbs reference
populations. \newrev{The corresponding secularized results are shown in Fig.~\ref{FIG1}(c)}}. Panels (b) and (c) show the real and imaginary parts of $\rho_{01}(t)$, respectively.}
        \label{coherency_population}
    \end{figure*}  
\subsection{Excited state initialization}
\label{sec3b}
Additionally, we provide in Appendix~\ref{MorePopDynamics} (Fig.~\ref{FIG12}) population dynamics for unchanged parameters; however, with the first excited state as initial state $\rho(0) = |1\rangle\langle 1|$, to observe vibrational relaxation. In that case, the first excited state population decreases while the ground state population increases until both reach the steady state depicted in Fig.~\ref{FIG1} (horizontal lines) for the given coupling strength. The relaxation times remain within the previously reported range 
 of equilibration times from several hundred femtoseconds to  picoseconds in the case of ground state initialization. This implies, in the context of vibrational relaxation in physical chemistry, that most of the coupling strengths presented here would be considered rather ``strong''~\cite{MayKuehn2011,Nitzan2006,Fischer2022}.

While the deviations of the mean-force Gibbs state to the regular Gibbs state remain unchanged, the Redfield equation clearly overestimates relaxation rates, especially for $\gamma \ge 1.0$. CCQME still delivers reasonable results compared to HEOM for $\gamma=1.0$, while stronger deviations 
 appear at $\gamma=5.0$}.

\subsection{Gaussian wavepacket initialization}
\label{sec3c}
In addition to initializing in an energy eigenstate, we also analyze the evolution of an initially left-localized wavepacket on the TAA double-well potential. \blue{The use of (Gaussian) wavepackets rather than system eigenstates as initial states is motivated by the fact that these describe more naturally reactive scattering processes, or initial conditions after laser pulse excitation~\cite{RamosFischer2024}. Further, a wavepacket as initial state has non-vanishing coherences (off-diagonal elements of the reduced density matrix), whose role will be briefly discussed below. }

\blue{
Specifically, we chose the system density matrix initially as} $\rho(0) = |\psi\rangle\langle \psi|$ with a normalized, moving Gaussian wavepacket,
\begin{equation}
\psi (q)
 =
\left(\frac{2}{\pi a^{2}}\right)^{1/4}
\exp\!\left[
-\frac{(q-q_{\mathrm L})^{2}}{a^{2}}
\right]
\exp\!\bigl\{ i k_{0}(q-q_{\mathrm L}) \bigr\},
\label{eq:init_wavepacket}
\end{equation}
where \(q_{\mathrm L}\) denotes the position of the minimum of the left well and \(a\) is the width parameter of the Gaussian. The wavepacket is assigned
a kinetic energy equal to the barrier height $E_{\mathrm B} = 1573.3 \ \text{cm}^{-1}$ of the double
well, such that
\(k_{0} = \sqrt{2 \mu_\text{TAA} E_{\mathrm B}}\).
The wavepacket is therefore initially centered at the minimum of
the left well, \(q_{\mathrm L}\), and the width parameter $a = 0.5 \ a_0$ is chosen
such that the packet is slightly narrower than the left well itself. Subsequently, the wavepacket begins moving toward the right. For this scenario we employ an extended twelve-level representation
(\(N = 12\)) of the TAA subsystem. 

Fig.~\ref{fig:wp_dyn_g05} shows the wavepacket dynamics at $\gamma = 0.5$. The left panel
displays the populations of the lowest four eigenstates \(\ket{0}\)–\(\ket{3}\),
while the right panel shows the corresponding expectation value of the
transfer coordinate \(\langle q\rangle(t)\). The latter shows new features, namely coherent oscillations, which depict \blue{the damped oscillation of the wavepacket in the asymmetric double well.}
These oscillations correspond to the mentioned non-zero off-diagonal elements in the reduced density matrix.
This is in contrast to the two previous initial states, where off-diagonal elements remain zero within the secular approximation. The decay of oscillations is faster for larger $\gamma$ and is \blue{analyzed in more detail} in Appendix~\ref{MoreWPDynamics} (Figs.~\ref{fig:wp_dyn_g00}-\ref{fig:wp_dyn_g10}). As $\gamma$ increases, the mean-force Gibbs state becomes important for an illustrative physical quantity, namely the steady-state value of the coordinate $q$, which is shifted compared to the regular Gibbs state. 
This is evident in Fig.~\ref{fig:wp_dyn_g05}(b), as well as in Fig.~\ref{fig:wp_dyn_g10}(b).

\blue{Moreover, we quantify the error of the transfer coordinate $\Delta q$ as a function of the coupling strength $\gamma$ in Fig.~\ref{fig:wp_dyn_bench} of Appendix~\ref{WPDynBench}, analogous to Fig.~\ref{fig:pop_error}. The errors are again computed as time averages up to $2.5$ ps. \newrev{For this wavepacket calculation, the secularized CCQME prediction for the transfer-coordinate expectation is closer to HEOM than the secularized Redfield prediction in this metric. This result should be viewed as a benchmark of the secularized prediction for $(\langle q\rangle(t))$, not as a general validation of the non-secular CCQME coherence dynamics for initially coherent states.}

\subsection{Limitations}
\label{sec3d}
\label{limit}

\blue{
The discussion above considered the secularized
implementations of Redfield and CCQME for the TAA proton-transfer model.
This approximation is expected to be well suited for the population dynamics (in particular if no coherences are present initially).}

\blue{To test this expectation and 
to assess the effect of population-coherence and coherence-coherence couplings which are missing in secular approximation, we consider here the non-secular Redfield and CCQME generators, as discussed in Sec.~\ref{Theory and formalism}, for the ground-state initial condition used in Fig.~\ref{FIG1}. The corresponding dynamics for $\gamma=1.0$ is shown in Fig.~\ref{coherency_population}. Panel (a) displays the populations of the three lowest energy eigenstates, while panels (b) and (c) show the real and imaginary parts of $\rho_{01}(t)$, respectively.
}

\blue{
The population dynamics in Fig.~\ref{coherency_population}(a)
shows that the performance of the non-secular CCQME is
state-dependent. For the {population of the} ground-state  $\ket{0}$,
the CCQME steady state is closer to the HEOM reference and to
the second-order mean-force Gibbs population than the Redfield
result. In contrast, for the first excited state $\ket{1}$, the
CCQME underestimates the HEOM and mean-force Gibbs populations,
whereas Redfield is closer for this particular component. For
the weakly populated state $\ket{2}$, Redfield relaxes toward a
negative population, while CCQME remains positive. Thus, the
non-secular calculation at the relatively strong coupling $\gamma=1.0$ does not provide
a uniform state-by-state improvement of CCQME over Redfield for populations. Note, however, that 
 CCQME avoids the unphysical, negative population of state $\ket{2}$.  }

\blue{
The limitations of CCQME are more pronounced in the coherence sector. As
shown in Fig.~\ref{coherency_population}(b), the real part of
$\rho_{01}(t)$ obtained from the non-secular CCQME does not relax
to the corresponding HEOM value and does not coincide with the
 second-order mean-force Gibbs coherence, given in Eq.~\eqref{eq:rho2_coherences_from_R2}, {while the} imaginary
part in Fig.~\ref{coherency_population}(c) shows a smaller
long-time deviation. }
%
\blue{Note, however, that through population-coherence transfer in the non-secular case,  errors incurred on the coherences are fed back 
into the populations, where they may introduce artificial features. This is manifested here by non-monotonic transient populations at around 100 fs {in Fig.~\ref{coherency_population}(a)}. These artefacts are not present in the secular approximation.
The deviation of $\mathrm{Re}[\rho_{01}(t)]$ in Fig.~\ref{coherency_population}(b) constitutes a
limitation of the present second-order non-secular application of
the CCQME, which is tailored to reproduce the second-order mean-force Gibbs state populations, {while the CCQME dissipator Eq.~\eqref{ct} does not produce second order corrected coherences. This} suggests that quantitatively reliable
coherence-sensitive observables require higher-order
finite-coupling corrections.
More details are presented in Appendix~\ref{RedfieldSO_derivation}}.

\section{Summary and Conclusions}
\label{conclusion}
We addressed intramolecular proton-transfer dynamics \blue{for a model representing} thioacetylacetone (TAA) in a solvent based on a Drude-cutoff Ohmic bath. The proton-transfer coordinate was modeled by an effective double-well potential and represented in a truncated vibrational eigenbasis of the renormalized system Hamiltonian $H_S$. We benchmarked the secularized Redfield  and canonically consistent quantum master equation (CCQME) against numerically exact hierarchical equations of motion (HEOM), focusing on both transient relaxation dynamics and steady-state populations over a broad range of coupling strengths $\gamma$. Our results demonstrate that, while the Redfield master equation is both dynamically and thermodynamically consistent for the TAA
subsystem in the weak coupling regime, \blue{it increasingly fails as the coupling strength increases to $\gamma=0.5$, thus entering what we classified here as the ``intermediate coupling'' regime.} In particular, Redfield dynamics relaxes to the Gibbs state regardless of the coupling strength, deviating from the HEOM reference. In contrast, the \newrev{secularized} CCQME consistently converges to the stationary populations closely following the mean-force Gibbs state up to second order in system-bath interaction. 

\newrev{As a result, for the population and steady-state
benchmarks, CCQME gives closer agreement with the HEOM} compared with the Redfield equation and extends the range
over which a second-order master-equation description remains quantitatively useful. For not too strong coupling  strengths, \newrev{the secularized CCQME also improves the transient population dynamics} compared to \newrev{the secularized} Redfield. \newrev{However, the representative non-secular calculation shows
that 
this improvement is not uniform once populations and coherences are coupled. More precisely, the non-secular result reveals a state-dependent performance: certain populations are not better reproduced by CCQME than by Redfield, while coherence-sensitive quantities exhibit noticeable deviations and short-time artefacts}, here for $\gamma = 1.0$ (Fig.~\ref{coherency_population}).
Therefore, the present results should be understood 
\newrev{as evidence for improved secularized steady-state population predictions}
within the
 (ultra)weak-to-intermediate coupling regime, rather than 
 \newrev{as a general improvement over Redfield across all couplings or beyond the secular approximation.}
We note, however, that no negative populations \newrev{were observed} for CCQME in contrast to Redfield, for the non-secular case studied in this work (Fig.~\ref{coherency_population}(a)). \blue{A recent polaron-transformed extension of the CCQME provides a promising route to access stronger system--bath coupling regimes while retaining the canonically consistent construction~\cite{Thingna2026}.}

In addition, we investigated the dynamics of a Gaussian wavepacket as initial state with kinetic energy equal to the barrier height, allowing us to apply the presented methods in a more complicated scenario, including coherent oscillations.} 

The present study establishes the \newrev{secularized} CCQME as a reliable and efficient tool for describing proton-transfer dynamics in multi-level molecular systems in the weak-coupling regime\newrev{, while the non-secular coherence sector requires further higher order corrections for quantitatively reliable predictions}. By providing a \newrev{secularized population and steady-state benchmark} against HEOM for a chemically relevant system, our study bridges the gap between developments of thermodynamically consistent master equations and their application in molecular quantum dynamics. The present framework paves the way for future investigations of solvent-controlled and cavity-modified proton-transfer reactions, where accurate steady-state properties and transient dynamics are  essential.

\section{Acknowledgment}
\label{Acknowledgment}
This work was funded by the Deutsche Forschungsgemeinschaft (DFG, German Research Foundation) –
CRC/SFB 1636 – Project ID 510943930 - Project No. A05.
\blue{JA gratefully acknowledges further support from the DFG, grants 513075417, 384846402.}

\appendix
\onecolumngrid
\newpage

\section{System energies and matrix elements of the coupling operator}
\label{tabels}
\section*{}

This appendix provides numerical data utilized in the present study. Table~\ref{tab:eigenvalues} lists the lowest six eigenenergies obtained by the effective system Hamiltonian $H_S$. Table~\ref{tab:matrixelements} provides the corresponding matrix elements of the proton-transfer coordinate operator $q$ in the same truncated energy eigenbasis of $H_S$ in the absence of system--bath coupling, such that $H_\text{ren}$ is set to zero.

\begin{table}[htbp]
\centering
\begin{tabular}{cc}
\hline
State $n$ & $E_n$  \\
\hline
0 & 4.114537$\times 10^{-3}$ \\
1 & 4.691015$\times 10^{-3}$ \\
2 & 8.133116$\times 10^{-3}$ \\
3 & 1.110714$\times 10^{-2}$ \\
4 & 1.458100$\times 10^{-2}$ \\
5 & 1.881039$\times 10^{-2}$ \\
\hline
\end{tabular}
\caption{Eigenvalues $E_n$ of the effective system Hamiltonian $H_S$, obtained from Eq.~\eqref{eq:eigen_HS} for the isolated system with $H_\text{ren}$  set to zero. The first six energy levels are shown in atomic units ($E_h$).}
\label{tab:eigenvalues}
\end{table} 

\begin{table}[htbp]
\centering
\begin{tabular}{c|cccccc}
\hline
$n\backslash m$ & 0 & 1 & 2 & 3 & 4 & 5 \\
\hline
0 &  -0.3813 &   0.3325 &   0.0837 &   0.1321 &   0.0564 &   0.0289 \\
1 &   0.3325 &   0.6712 &  -0.2931 &   0.0230 &  -0.0498 &   0.0008 \\
2 &   0.0837 &  -0.2931 &   0.4089 &  -0.4241 &   0.0559 &  -0.0514 \\
3 &   0.1321 &   0.0230 &  -0.4241 &   0.1598 &  -0.5011 &  -0.0085 \\
4 &   0.0564 &  -0.0498 &   0.0559 &  -0.5011 &   0.2696 &  -0.5258 \\
5 &   0.0289 &   0.0008 &  -0.0514 &  -0.0085 &  -0.5258 &   0.2752 \\
\hline
\end{tabular}
\caption{Matrix elements $q_{nm}=\langle n|q|m\rangle$ of the proton-transfer coordinate operator for the TAA system (Eq.~\eqref{eq:q_operator}), evaluated in the energy eigenbasis of $H_{S}$ with $H_\text{ren}$  set to zero. Values are shown for the lowest six eigenstates in atomic units ($a_0$).}
\label{tab:matrixelements}
\end{table}
    \section{Temporal Correlations in a Thermal Bath}\label{TemporalCorrelationsBath}
\section*{}
\blue{This appendix reviews the standard construction of thermal-bath
correlation functions and the Ohmic-Drude spectral density used in
open-quantum-system theory and quantum dissipation
\cite{Kubo1991,Mukamel1995,BreuerPetruccione2007,Weiss2012,
Leggett1987,CaldeiraLeggett1983,Grabert1988,CallenWelton1951}.
 The material
in this appendix is included for completeness and reproducibility.}

In the framework of open quantum systems, the bath significantly affects the evolution of the system's states due to its continuous interaction with the system. At any specific time \( t \), the state of both the system and the bath in the Schrödinger picture can be expressed as
\begin{equation}
\rho_{\text{tot}}(t) =
   U_{\text{tot}}(t) \, \rho_{\text{tot}}(0)\,
U_{\text{tot}}^\dagger(t),\qquad U_{\text{tot}}(t) =\exp\!\big[-i(H_S + H_B + H_{SB})t\big],
\label{interaction_picture_coupling_operator}
\end{equation}
where
$U_{\text{tot}}(t) $
denotes the unitary evolution operator of the total Hamiltonian. The Hamiltonian of $H_S$, $H_B$ and $H_{SB}$ are given in Eq.~\eqref{eq:HS}, Eq.~\eqref{HB} and Eq.~\eqref{eq:HSB}, respectively.  To derive the dynamics of the system, we trace over the bath degrees of freedom
\begin{equation}
   \rho(t) = \mathrm{Tr}_B\{\rho_{\text{tot}}(t)\},
   \label{rho_definition}
\end{equation}
showing that the specific details of the bath correlations are directly incorporated into the reduced system
state. The bath operator (Eq.~\eqref{eq:X_operator}) evolves in the Heisenberg picture with respect to $H_B$ 
\begin{equation}
\mathbb{X}(t)=e^{iH_B t}\,\mathbb{X}\,e^{-iH_B t}.
\label{11}
\end{equation}
\indent Bath correlation functions characterize the bath’s memory by quantifying how strongly bath fluctuations at time 
0
remain correlated with those at a later time 
$t$. The two point bath correlation function in the interaction Hamiltonian $H_{SB}$ is \blue{expressed as \cite{Kubo1991, Mukamel1995, Ishizaki2009}}

\begin{equation}
C_F(t)=\langle \mathbb{X}(t)\mathbb{X}(0)\rangle_B,
\label{correlation_function}
\end{equation}
where $\langle \cdot \rangle_B=\mathrm{Tr}_B(\rho_B\,\cdot)$ denotes the expectation value in the thermal bath state $\rho_B$. When the correlations diminish rapidly, the bath quickly forgets its past states (Markovian regime). Conversely, if the correlations decay slowly, the bath retains its memory resulting in
non-Markovian effects.
Using Eq.~\eqref{eq:X_operator}, the Heisenberg operator in Eq.~(\ref{11}) can be rewritten as
\begin{equation}
\mathbb{X}(t) = - \sum_k c_k \sqrt{\frac{1}{2 m_k \omega_k}} \ \big(b_k(t)+ b_k^\dagger (t)\big),
\label{Heisenberg-picture}
\end{equation}
where $c_k$ is the coupling constant. In atomic units $\hbar=1$ is omitted.
Here, $b_k^\dagger(t)=b_k^\dagger e^{i \omega_k t}$ and $b_k(t) = b_k e^{-i \omega_k t}$ are the bosonic creation and annihilation operators, respectively, evolving in the Heisenberg picture.
Therefore Eq.~\eqref{correlation_function} reads

\begin{equation}
C_F(t)=\left\langle \sum_{j,k}\frac{c_j c_k}{2\sqrt{m_j m_k\,\omega_j\omega_k}}
\,(b_j e^{-i\omega_j t}+b_j^\dagger e^{i\omega_j t})(b_k+b_k^\dagger)\right\rangle_B.
\end{equation}
Cross terms with $j \neq k$ vanish, while those with $j = k$ survive. Moreover, the expectation values $\langle b_k b_k \rangle = \langle b_k^\dagger b_k^\dagger \rangle = 0$. For each mode $k$
\[
\langle b_k^\dagger b_k\rangle = n(\omega_k)=\frac{1}{e^{\beta\omega_k}-1},
\quad
\langle b_k b_k^\dagger\rangle = 1+n(\omega_k),
\]
where $\beta =1/(k_BT)$ ($k_B=1$ is the Boltzmann constant and $T$ the temperature). Thus
\begin{equation}
C_F(t)
=\sum_k \,\frac{c_k^2}{2m_k\omega_k}
\Bigl[(1+n(\omega_k))\,e^{-i\omega_kt}+n(\omega_k)\,e^{i\omega_kt}\Bigr].
\end{equation}
The spectral density is defined as
\begin{equation}
\mathcal{J}'(\omega)
=\frac{\pi}{2}\sum_k\frac{c_k^2}{m_k\,\omega_k}\,\delta(\omega-\omega_k)\,.
\end{equation}
Using the identity
\begin{equation}
\sum_k\frac{c_k^2}{2m_k\omega_k}\,f(\omega_k)
\;=\;
\frac{1}{\pi}\int_{0}^{\infty}\!d\omega\;\mathcal{J}'(\omega)\,f(\omega),
\end{equation}
the bath correlation function can be expressed in Fourier form as
\begin{equation}
C_F(t)
=\frac{1}{\pi}
\int_{0}^{\infty}d\omega\;\mathcal{J}'(\omega)\;
\Bigl[(1+n(\omega))\,e^{-i\omega t}+n(\omega)\,e^{i\omega t}\Bigr],
\end{equation}
where $\omega$ is the frequency.

Direct evaluation of the oscillatory integrals is generally challenging.
A convenient simplification uses the Bose function identity $~1+n(\omega) = -\,n(-\omega)$
allowing one to introduce a frequency-symmetric continuation of the spectral density.
Defining
\begin{equation}
\mathcal{J}(\omega) \;\equiv\; \mathcal{J}'(\omega)\,\Theta(\omega)\;-\;\mathcal{J}'(-\omega)\,\Theta(-\omega),
\label{eq:J_symmetric}
\end{equation}
with $\Theta(\omega)$ the Heaviside step function, one obtains~\cite{ BeckerDiss2022}
\begin{equation}
\mathcal{J}(\omega) =
\begin{cases}
\mathcal{J}'(\omega), & \omega > 0, \\[6pt]
0, & \omega = 0, \\[6pt]
-\,\mathcal{J}'(-\omega), & \omega < 0,
\end{cases}
\label{step_function}
\end{equation}
which is odd in $\omega$. In this way, the correlation function can be expressed as a single Fourier transform
\begin{equation}
C_F(t)
= \int_{-\infty}^{\infty} \frac{d\omega}{\pi}\; e^{i\omega t}\, \mathcal{J}(\omega)\,n(\omega).
\label{eq:CF_symmetric}
\end{equation}
This representation removes the splitting of positive and negative frequencies. \blue{A detailed discussion for
spectral densities is given in Ref.~\citenum{Weiss2012}.}

\blue{The contour-integration evaluation below is a standard derivation of
the Matsubara expansion of the Ohmic--Drude bath correlation function. Closely related derivations are given in 
Refs.~\citenum{BeckerDiss2022, Weiss2012,RitschelEisfeld2014}.} Using the symmetry of the Bose-Einstein distribution $n(\omega)$, one may equivalently extend the integration to the entire real axis defining the integrand as
\begin{equation}
g(\omega;t) \equiv e^{i\omega t} \mathcal{J}(\omega)\, n(\omega).
\end{equation}
For $t>0$, the exponential factor $e^{i\omega t}$ suppresses all contributions with $\Im(\omega) > 0$. Therefore, by closing the contour in the upper
half-plane and applying Cauchy’s theorem, the original integral can be expressed as a sum over the poles
of $g(\omega;t)$ placing inside the contour. These poles originate from two distinct sources. To be more exact, 
the first contribution comes from the spectral density. Using the Ohmic–Drude form
\begin{equation}
\mathcal{J}(\omega) =  \gamma \, \frac{\omega}{1 + \omega^2/\omega_c^2}
= \gamma \, \frac{\omega_c^2 \, \omega}{\omega^2 + \omega_c^2},
\label{25}
\end{equation}
one finds that it has a simple pole in the upper half-plane at $\omega = i \omega_c$, resulting from
the analytic structure of the Drude cutoff function. The second contribution arises from the Bose–Einstein distribution, having simple poles along the
imaginary axis at the Matsubara frequencies
\begin{equation}
\omega = i \nu_n,
\qquad \nu_n = \frac{2 \pi n}{\beta},
\qquad n = 1,2,\dots,
\label{26}
\end{equation}
with corresponding residues
$
\mathrm{Res}\big[n(\omega), \omega = i \nu_n\big] = \dfrac{1}{\beta}.$ These poles account for the thermal contributions of the bath.The residue of the spectral density at the Drude pole is
\begin{equation}
\mathrm{Res}[\,\mathcal{J}(\omega),\,\omega=i\omega_c\,]
= \lim_{\omega\to i\omega_c} (\omega-i\omega_c)\mathcal{J}(\omega)
= \frac{\gamma \,\omega_c^2}{2}.
\label{eq:Res_Drude}
\end{equation}
For the Matsubara contributions, only the values of $\mathcal{J}(\omega)$ on the imaginary axis are required
\begin{equation}
\mathcal{J}(i\nu_n) = \gamma\;\frac{i\nu_n}{1-\nu_n^2/\omega_c^2}.
\label{eq:J_Matsubara}
\end{equation}
Collecting all contributions, the correlation function becomes
\begin{equation}
C_F(t) =
2\,\mathrm{Res}[\,\mathcal{J}(\omega),\,\omega=i\omega_c\,]\; i\,n(i\omega_c)\,e^{-\omega_c t}
+ \frac{2}{\beta}\sum_{n=1}^{\infty} i\,\mathcal{J}(i\nu_n)\,e^{-\nu_n t}.
\label{eq:CF_general_Matsubara}
\end{equation}
Substituting Eqs.~\eqref{eq:Res_Drude} and \eqref{eq:J_Matsubara}, one finds the explicit form
\begin{equation}
C_F(t)=
i\,(\gamma\,\omega_c^2)\,n(i\omega_c)\,e^{-\omega_c t}
- \frac{2 \gamma}{\beta}\sum_{n=1}^{\infty}\frac{\nu_n}{1-\nu_n^2/\omega_c^2}\,e^{-\nu_n t}.
\label{eq:CF_final}
\end{equation}
Writing the Bose function in terms of the hyperbolic cotangent
$n(\omega) = \left(\coth(\beta\omega/2) - 1\right)/2,$
Eq.~\eqref{eq:CF_final} can be rewritten as follows:
\begin{equation}
C_F(t)= \frac{ \gamma \omega_{c}^{2}}{2}
\left[\cot\left(\frac{\beta \omega_{c}}{2}\right) - i \right] e^{-\omega_{c}t}
- \frac{2 \gamma}{\beta}\sum_{n=1}^{\infty}\frac{\nu_n}{1-\nu_n^2/\omega_c^2}\,e^{-\nu_n t} .
\label{matsubara_tang}
\end{equation}
Eq.~\eqref{matsubara_tang} displays the correlation function as the sum of two qualitatively different parts:
(i) the single Drude pole term, decaying on the bath memory scale $\omega_c^{-1}$, and (ii) an infinite series of Matsubara exponentials, each controlled by thermal frequencies $\nu_n = 2\pi n/\beta$.
The first captures the intrinsic cutoff of the spectral density, while the second accounts for the thermal occupation of the bath modes. This representation signifies that the bath memory is generally very short at high temperatures or large cutoffs, providing justification for Markovian approximations.

\section{Temperature-dependent tunneling rate $\mathbb{T}(\delta)$}
\label{TemperaturedependentTunneling}
\section*{}

In the interaction picture with respect to free Hamiltonian
$(H_S+H_B)$, the system operator $q$ (Eq.~\eqref{eq:q_operator}) evolves only with the system Hamiltonian $H_S$, thus
\begin{equation}
\tilde{q}(t) = e^{iH_S t}\, q\, e^{-iH_S t}.
\end{equation}
For a backward time argument ($-t$), one finds
\begin{equation}
\tilde{q}(-t)= e^{-iH_S t}\, q\, e^{iH_S t}.
\end{equation}
Taking matrix elements in the $\{|n\rangle\}$ basis gives
\begin{equation}
\langle n| \tilde{q}(-t)|m\rangle
    = e^{-i(E_n - E_m) t}\, q_{nm}
    \equiv e^{-i\delta_{nm} t}\, q_{nm},
\label{bohr_frequency}
\end{equation}
where $\delta_{nm}=E_n-E_m$ is the Bohr frequency.

Within a perturbative treatment up to second order in the interaction Hamiltonian
$H_{SB}$, one arrives at the Redfield master equation, with the Redfield superoperator given in Eq.~\eqref{eq:R2_def} (Appendix~\ref{PPerturbative master equations} provides more details). In this formulation, the bath correlation function $C_F(t')$ enters via the convolution  ~\cite{BeckerDiss2022} 
\begin{equation}
\mathcal{K}_t = \int_0^t C_F(t')\,\tilde{q}(-t')\, d{t}'.
\label{convoluted}
\end{equation}
After expressing the convolution operator $\mathcal{K}_t$ in this eigenbasis, the matrix elements take the form
\begin{equation}
\langle n| \mathcal K_t |m\rangle
    = \int_0^t C_F(t')\,
       \langle n| \tilde{q}(-t')|m\rangle\, d{t}' 
    = \mathbb{T}_{t}(\delta_{nm})\, q_{nm},
    \label{convoluted_1}
\end{equation}
where
\begin{equation}
\mathbb{T}_{t}(\delta_{nm}) = \int_0^t e^{-i \delta_{nm} t'} \, C_F(t')\, d{t}',
\label{tunneling_rate}
\end{equation}
defines the time-dependent tunneling rate. In the asymptotic limit $t\to\infty$ (Markovian regime), one obtains
\begin{equation}
\mathbb{T}(\delta_{nm}) = \lim_{t \to \infty} \mathbb{T}_{t}(\delta_{nm}),
\label{tunneling_rate_2}
\end{equation}
which is the half-sided Fourier transform of the bath correlation function.
Substituting the Fourier representation of $C_F(t)$ \cite{Redfield1965,BreuerPetruccione2007,Weiss2012,MayKuehn2011,Nitzan2006} from Eq.~\eqref{eq:CF_symmetric}, one finds
\begin{equation}
\mathbb{T}(\delta_{nm})\;=\; \mathcal{J}(\delta_{nm})\, n_\beta(\delta_{nm}) \;+\; i \, \mathcal{P}\!\int_{-\infty}^{\infty}
    \frac{\mathcal{J}(\omega)\, n_\beta(\omega)}{\omega - \delta_{nm}}\, \frac{d\omega}{\pi},
    \label{ttunneling_rate}
\end{equation}
where $\mathcal{J}(\omega)$ is the spectral density of the bath and
$n_\beta(\omega) = (e^{\beta \omega}-1)^{-1}$ is the Bose–Einstein occupation factor. The first term
\begin{equation}
\mathcal{J}(\delta_{nm})\, n_\beta(\delta_{nm})=\Re{\big[\mathbb{T}(\delta_{nm})\big]} ,
\label{real_tunneling_rate}
\end{equation}
is real and determines the transition rates induced by the bath, and the second term
\begin{equation}
\mathcal{P}\!\int_{-\infty}^{\infty}    \frac{\mathcal{J}(\omega)\, n_\beta(\omega)}{\pi(\omega - \delta_{nm})}\, d\omega=\Im{\big[\mathbb{T}(\delta_{nm})\big]} ,
\label{imaginary_tunneling_rate}
\end{equation}
is purely imaginary, giving rise to coherent level shifts (Lamb shifts).

To evaluate the canonical tensor in Eq.~\eqref{ct}, we require the first derivative of the tunneling rate defined in Eq.~\eqref{tunneling_rate}. It can be written in terms of the Drude pole and Matsubara series, explained in Appendix~\ref{TemporalCorrelationsBath}. Inserting Eq.~\eqref{matsubara_tang} into Eq.~\eqref{tunneling_rate} gives
\begin{equation}
\mathbb{T}_{t}(\delta_{nm})=\int_0^t\frac{\gamma\,\omega_c^{2}}{2}\,
\Bigg[\cot\!\Big(\frac{\beta \omega_c}{2}\Big)-i\Bigg]\,
e^{-(\omega_c+i\delta_{nm})t'}\,dt'
-\int_0^t \sum_{n=1}^{\infty}
\Bigg[\frac{2\gamma}{\beta}\,
\frac{\nu_n}{1-(\nu_n/\omega_c)^2}\Bigg]\,
e^{-(\nu_n+i\delta_{nm})t'}\,dt', 
\label{tunneling_matsubara}
\end{equation}
 where ${\nu}_{n}$ is given in Eq.~\eqref{26}. For $a\in\{\omega_c,\nu_n\}$ with $a>0$, one can use the following identity
\begin{equation}
\int_0^t e^{-(a+i\delta_{nm})t'}\,dt'=\frac{1-e^{-(a+i\delta_{nm})t}}{a+i\delta_{nm}}.
\end{equation}
Therefore, Eq.~\eqref{tunneling_matsubara} can be rewritten as
\begin{equation}
\mathbb{T}_t(\delta_{nm})=\frac{\gamma\,\omega_c^{2}}{2}\,
\Bigg[\cot\!\Big(\frac{\beta \omega_c}{2}\Big)-i\Bigg]\,
\frac{1-e^{-(\omega_c+i\delta_{nm})t}}{\omega_c+i\delta_{nm}} - \sum_{n=1}^{\infty}
\Bigg[\frac{2\gamma}{\beta}\,
\frac{\nu_n}{1-(\nu_n/\omega_c)^2}\Bigg]\,
\frac{1-e^{-(\nu_n+i\delta_{nm})t}}{\nu_n+i\delta_{nm}}\,.
\label{tunneling-rate-3}
\end{equation}
When $t\to\infty$, the transients vanish. Thus
\begin{equation}
\mathbb{T}(\delta_{nm})=\frac{A}{\omega_c+i\delta_{nm}}
+\sum_{n=1}^{\infty}\frac{B_n}{\nu_n+i\delta_{nm}}\,,
\label{tunnelingrate_matsubara}
\end{equation}
where 
\begin{equation}
A=\frac{\gamma\,\omega_c^{2}}{2}\Bigg[\cot\!\Big(\frac{\beta \omega_c}{2}\Big)-i\Bigg],
\qquad
B_n=-\frac{2\gamma}{\beta}\,\frac{\nu_n}{1-(\nu_n/\omega_c)^2}\,.
\end{equation}
After some algebra the first derivative of the tunneling rate reads

\begin{align}
\frac{\partial}{\partial \delta_{nm}}\big[\mathbb{T}(\delta_{nm})\big]
&= -\,i\,\frac{A}{(\omega_c+i\delta_{nm})^2}
\;-\;\sum_{n=1}^{\infty} i\,\frac{B_n}{(\nu_n+i\delta_{nm})^2}\nonumber\\
&= -\,i\,\frac{\gamma\,\omega_c^{2}}{2}
\Bigg[\cot\!\Big(\frac{\beta \omega_c}{2}\Big)-i\Bigg]\frac{1}{(\omega_c+i\delta_{nm})^2}
\;+\; i\sum_{n=1}^{\infty}\frac{2\gamma}{\beta}\,
\frac{\nu_n}{1-(\nu_n/\omega_c)^2}\,
\frac{1}{(\nu_n+i\delta_{nm})^2}.
\label{derivetive_matsubara}
\end{align}

\section{Perturbative master equation}
\label{PPerturbative master equations}
\section*{}
\blue{This section reviews the derivation of the Redfield and CCQME equation using a second-order perturbative expansion in the system-bath coupling $(H_{SB})$. Derivations can also be found, e.g., in
Refs.~\citenum{Carmichael1999,Weiss2012,BreuerPetruccione2007,CohenTannoudji1998, GardinerZoller2004, WallsMilburn2008,Becker2022, BeckerDiss2022}}.
In the weak-coupling regime, the influence of the environment on the system can be addressed using perturbative methods applied to the system-bath interaction, $H
_{SB}$. Initially, the system and bath are assumed to be uncorrelated

\begin{equation}
\rho_{\mathrm{tot}}(0)=\rho(0)\otimes \rho_B,
\label{eq:factorized_initial}
\end{equation}
where $\rho_B=e^{-\beta H_B}/Z_B$ is the thermal equilibrium state of the bath and $[H_B,\rho_B]=0$. The total density matrix in the interaction picture with respect to the free Hamiltonian is defined as
\begin{equation}
\tilde{{\rho}}_{\mathrm{tot}}(t)=e^{i(H_S+H_B)t}\,\rho_{\mathrm{tot}}(t)\,e^{-i(H_S+H_B) t},
\label{totoal_density_interaction_picture}
\end{equation}
where $\rho_{tot}(t)$ is given in Eq.~\eqref{interaction_picture_coupling_operator}. According to the Von Neumann equation

\begin{equation}
\frac{\partial \tilde{\rho}_{\mathrm{tot}}(t)}{\partial t}
=
-i\,\big[\tilde{H}_{SB}(t),\tilde{\rho}_{\mathrm{tot}}(t)\big],
\label{interaction-picture}
\end{equation}
where $\tilde{H}_{SB}(t)=e^{i(H_S+
H_B)t} H_{SB} e^{-i(H_S+
H_B)t}$ is the interaction Hamiltonian in the interaction picture. Integrating Eq.~\eqref{interaction-picture} and expanding the time-ordered exponential (Dyson series)~\cite{BeckerDiss2022} up to second order in 
$H_
{SB}$
gives

\begin{equation}
\tilde{\rho}_{\mathrm{tot}}(t)
\simeq
\tilde{\rho}_{\mathrm{tot}}(0)
-i\int_{0}^{t}\!dt'_1\,[\tilde{H}_{SB}(t'_1),\tilde{\rho}_{\mathrm{tot}}(0)]
-\int_{0}^{t}\!dt'_1\int_{0}^{t'_1}\!dt'_2\,[\tilde{H}_{SB}(t'_1),[\tilde{H}_{SB}(t'_2),\tilde{\rho}_{\mathrm{tot}}(0)]].
\label{eq:appB_Dyson_tot}
\end{equation}
The reduced density matrix is obtained by taking the partial trace over the bath's degrees of freedom
\begin{equation}
\tilde{\rho}(t)
\simeq
\rho(0)
-i\int_{0}^{t}\!dt'_1\,\tr_B\!\Big([\tilde{H}_{SB}(t'_1),\rho(0)\!\otimes\!\rho_B]\Big)
-\int_{0}^{t}\!dt'_1\int_{0}^{t'_1}\!dt'_2\,\tr_B\!\Big([\tilde{H}_{SB}(t'_1),[\tilde{H}_{SB}(t'_2),\rho(0)\!\otimes\!\rho_B]]\Big).
\label{eq:appB_Dyson_red}
\end{equation}
Because of symmetry, the odd orders vanish, and only even orders contribute to the Dyson expansion above.
Thus, Eq.~\eqref{eq:appB_Dyson_red}  defines a reduced dynamical map for the interaction-picture state  
$
{\rho}(0)\ \mapsto\ \tilde{\rho}(t)=e^{iH_S t}\rho(t)e^{-iH_S t}$,  
expressed as a perturbative series in the  Hamiltonian $H_{SB}$.
 A standard approach to achieve the evolution equation is to differentiate the truncated map (Eq.~\eqref{eq:appB_Dyson_red}). Returning to the Schrödinger picture $\rho(t)=e^{-iH_S t}\tilde{\rho}(t)e^{iH_S t}$, one \blue{arrives at ~\cite{Breuer2016}}

\begin{equation}
\frac{\partial\rho(t)}{\partial t}
\simeq
-i[H_S,\rho(t)]
+\mathcal{R}^{(2)}_t\!\Big[\,e^{-iH_S t}\rho(0)e^{iH_S t}\Big],
\label{eq:appB_nonlocal_initial}
\end{equation}
where $\mathcal{R}^{(2)}_t$ is a second-order time-dependent Redfield superoperator defined as 

\begin{equation}
\mathcal{R}^{(2)}_{t}\!\Big[e^{-iH_S t}\,\rho(0)\,e^{iH_S t}\Big]
=
-\int_{0}^{t}\! dt'\;
\tr_B\!\Big[
H_{SB},\,
\Big[
\tilde{H}_{SB}(-t'),\,
e^{-iH_S t}\,\rho(0)\,e^{iH_S t}\otimes\rho_B
\Big]
\Big].
\label{eq:R2_doublecomm}
\end{equation}
 The Redfield term still depends on the initial state $\rho(0)$, so the evolution is not expressed solely in terms of the current state $\rho(t)$. To obtain a closed equation in terms of $\rho(t)$, one can approximate the inverse map to lowest order in the interaction by the free unitary evolution
\begin{equation}
\rho(0)\ \simeq\ e^{iH_S t}\,\rho(t)\,e^{-iH_S t},
\label{eq:appB_inverse_approx}
\end{equation}
producing the standard time-local Redfield master equation 

\begin{equation}
\frac{\partial\rho(t)}{\partial t}
\simeq
-i[H_S,\rho(t)]+\mathcal{R}^{(2)}_t[\rho(t)].
\label{eq:appBB_redfield_tlocal}
\end{equation}
After expanding the two commutators present in Eq.~\eqref{eq:R2_doublecomm}, the second-order Redfield superoperator for a Hermitian coupling operator $q = q^{\dagger}
$ can be expressed as follows:

\begin{align}
\mathcal{R}^{(2)}_{t}[\rho(t)]
&=
\int_{0}^{t}\! dt'\;
\tr_B\!\Big[
\tilde{H}_{SB}(-t')\,\rho(t)\otimes\rho_B\,H_{SB}
-
H_{SB}\,\tilde{H}_{SB}(-t')\,\rho(t)\otimes\rho_B
\Big]
+\mathrm{H.c.}\\
&=\mathcal{K}_t\,\rho(t)\,q
- q\,\mathcal{K}_t\,\rho(t)
+ q\,\rho(t)\,\mathcal{K}_t^\dagger
- \rho(t)\,\mathcal{K}_t^\dagger q,
\label{eq:R2_def_time-dependent}
\end{align}
where ${\mathcal{K}}_t$, $\rho(t)$ and $q$ are represented in Eq.~\eqref{convoluted}, Eq.~\eqref{rho_definition} and Eq.~\eqref{eq:q_operator}, respectively. 
 In the Markovian limit, the convolution operator (Eq.~\eqref{convoluted}) reads
\begin{equation}
\mathcal{K}=\lim_{t\to\infty}\mathcal{K}_t
=
\int_0^{\infty}\!dt'\, C_F(t')\,\tilde{q}(-t'),
\label{eq:appB_K_def}
\end{equation}
and the corresponding asymptotic Redfield generator
\begin{equation}
\mathcal{R}^{(2)}[\rho(t)]
=
\mathcal{K}\,\rho(t)\,q
- q\,\mathcal{K}\,\rho(t)
+ q\,\rho(t)\,\mathcal{K}^\dagger
- \rho(t)\,\mathcal{K}^\dagger q.
\label{eq:appB_R_def}
\end{equation}

The second-order Redfield superoperator collects the second-order contribution in the Dyson expansion of the reduced dynamical map for the interaction-picture state, so that it relates $\tilde{\rho}(t)=e^{iH_S t}\rho(t)e^{-iH_S t}$ to $\rho(0)$ after tracing out the bath.
Assuming that the zeroth-order evolution is given by the coherent propagation under $H_S$,
one can write the second-order approximation in the form
\begin{equation}
\rho(t)\ \simeq\ \bigl(1+\mathcal C_t^{(2)}\bigr)\!\Big[e^{-iH_S t}\,\rho(0)\,e^{iH_S t}\Big],
\label{dddd}
\end{equation}
where $\mathcal C_t^{(2)}$ represents the $\mathcal O(H_{SB}^{(2)})$ correction to the reduced dynamical map.
The map $\mathcal C_t^{(2)}$ is identified from the truncated Dyson expansion in Eq.~\eqref{eq:appB_Dyson_red}
\begin{equation}
\tilde{\rho}(t)\ \simeq\ \rho(0)+\tilde{\mathcal C}_t^{(2)}[\rho(0)],
\qquad
\tilde{\mathcal C}_t^{(2)}[\rho]
=
-\int_{0}^{t}\!dt'_1\int_{0}^{t'_1}\!dt'_2\,
\tr_B\!\Big([\tilde{H}_{SB}(t'_1),[\tilde{H}_{SB}(t'_2),\rho\otimes\rho_B]]\Big),
\label{eq:Ct2_def_from_Dyson}
\end{equation}
and in the Schr\"odinger picture yields

\begin{equation}
\mathcal C_t^{(2)}
=
e^{-i[H_S,\cdot]\,t}\,\tilde{\mathcal C}_t^{(2)}\,e^{+i[H_S,\cdot]\,t}
=
\int_0^{t}\! dt'\;
e^{-i[H_S,\cdot]\,(t-t')}\,\mathcal R_{t'}^{(2)}\,e^{+i[H_S,\cdot]\,(t-t')},
\label{eq:Ct2_def_integrated}
\end{equation}
which is consistent with Eq.~\eqref{dddd} up to $\mathcal O(H_{SB}^2)$,  signifying that $\mathcal C_t^{(2)}$ is built from second-order dynamical information only.

\blue{The CCQME construction used below is not a new theoretical approach
developed in the present work. It follows the canonically consistent
quantum master equation introduced in
Refs.~\citenum{Becker2022,BeckerDiss2022}. }

Using the formal inverse of the truncated map $(1+\mathcal C_t^{(2)})^{-1}\simeq 1-\mathcal C_t^{(2)}$
and neglecting $\mathcal O(H_{SB}^4)$, one finds
\begin{equation}
e^{-iH_St}\,\rho(0)\,e^{iH_St}
\simeq
(1-\mathcal C_t^{(2)})[\rho(t)].
\label{eq:free_state_via_inverse_map}
\end{equation}
Inserting Eq.~\eqref{eq:free_state_via_inverse_map} into Eq.~\eqref{eq:appB_nonlocal_initial}
leads to the canonically consistent quantum master equation
\begin{equation}
{\frac{{\partial}}{\partial t}}\big[{\rho{(t)}}\big]_\text{CCQME}
=
-i[H_S,\rho(t)]
+
\mathcal R_t^{(2)}\!\Big[(1-\mathcal C_t^{(2)})[\rho(t)]\Big].
\label{ccqqmmee}
\end{equation}
\blue{The steady state of Eq.~\eqref{ccqqmmee} is now set to be the (second-order)} mean-force Gibbs state, i.e. 
\begin{equation}
\big[\rho(t\to\infty)\big]_\text{CCQME}=\tau_{\mathrm{MF}}^{(2)}
=
\tau_{\mathrm{G}}
+\underbrace{
\mathcal C^{(2)}[\tau_{\mathrm{G}}]
-
\tau_{\mathrm{G}}\,\tr_S\!\left(\mathcal C^{(2)}[\tau_{\mathrm{G}}]\right)}_{\Delta\tau_{\mathrm{MF}}^{(2)}},
\label{eq:rhoMFviaQ2}
\end{equation}
where $\mathcal C^{(2)}[\tau_{\mathrm{G}}]=\tau^{(2)}$ represents the unnormalized second-order contribution, while the subtraction term ensures normalization of the mean-force Gibbs state at order $H_{SB}^{(2)}$.

\par
To verify that Eq.~\eqref{ccqqmmee} relaxes to the mean-force Gibbs state up to second order, one can take the asymptotic limit $\mathcal R_t^{(2)}\to \mathcal R^{(2)}$ and $\mathcal C_t^{(2)}\to \mathcal C^{(2)}$, and evaluate it on the unnormalized second-order equilibrium  
$\tau_{\mathrm G}+\mathcal C^{(2)}[\tau_{\mathrm G}]$ . 
Hence~\cite{Becker2022, BeckerDiss2022},
\begin{equation}
\begin{aligned}
\frac{\partial}{\partial t}\,(1+\mathcal C^{(2)})[\tau_G]
&=
-i[H_S,(1+\mathcal C^{(2)})[\tau_G]] \\
&\quad
+\mathcal R^{(2)}\!\left[(1-\mathcal C^{(2)})(1+\mathcal C^{(2)})[\tau_{\mathrm G}]\right].
\end{aligned}
\label{eq:ccqme_validity_step1}
\end{equation}
Using $[H_S,\tau_{\mathrm G}]=0$ and the second-order approximation $(1-\mathcal C^{(2)})(1+\mathcal C^{(2)}) \simeq 1$, one arrives at
\begin{equation}
\frac{\partial}{\partial t}\,(1+\mathcal C^{(2)})[\tau_G]
\simeq
-i[H_S,\tau^{(2)}]+\mathcal R^{(2)}[\tau_{\mathrm G}] = 0.
\label{eq:ccqme_validity_step3}
\end{equation}
\blue{The last equality comes from the equilibrium condition discussed in Appendix~\ref{RedfieldSO_derivation} below, see Eq.~\eqref{eq:stationarity_order22}.} \blue{Thus, the deviation observed between the CCQME steady state and the second-order mean-force Gibbs state at $\gamma=5.0$ in Fig.~\ref{FIG1}(d) 
indicates that the higher-order finite-coupling correction represented by the neglected term $(\mathcal C^{(2)})^2$ is no longer negligible at such coupling strengths.}

\onecolumngrid
\section{Equilibrium steady state}
\label{RedfieldSO_derivation}
\section*{}

The exact time-local master equation in the weak-coupling regime can be expanded perturbatively in the system--bath interaction $H_{SB}$ as
\begin{equation}
\frac{\partial\rho(t)}{\partial t}
=
-i[H_S,\rho(t)] + \mathcal R^{(2)}[\rho(t)] + \mathcal R^{(4)}[\rho(t)] + \cdots,
\label{eq:ME_generator_expansion}
\end{equation}
where $\mathcal R^{(2)}[\rho(t)]$ denotes the second-order Redfield superoperator given in Eq.~\eqref{eq:appB_R_def} and $\mathcal R^{(4)}[\rho(t)]$ collects the fourth-order contribution to the time-local generator. In the asymptotic limit, $
\left.\frac{\partial \rho(t)}{\partial t}\right|_{t\to\infty}=0
$, a stationary state satisfies
\begin{equation}
0= -i[H_S,\tau_{MF}] + \mathcal R^{(2)}[\tau_{MF}] + \mathcal R^{(4)}[\tau_{MF}] + \cdots,
\label{eq:stationarity_exact_hierarchy}
\end{equation}
where
\begin{equation}
\tau_{MF}=\tau_G+{\Delta\tau_{\mathrm{MF}}^{(2)}}+\mathcal O(H_{SB}^4),
\label{eq:rho_infty_seriess}
\end{equation}
is the mean-force Gibbs state. Here $\tau_{\mathrm G}$ is the Gibbs state given in Eq.~\eqref{eq:rho0}, and the second-order correction is ~\cite{BeckerDiss2022,Cresser2021}
\begin{equation}
\Delta\tau_{\mathrm{MF}}^{(2)}
=
\tau^{(2)}-\tau_n^{(2)},
\label{eq:rho_RG2}
\end{equation}
with
\begin{equation}
\tau^{(2)}
=
\tau_{\mathrm G}\,\tr_{B}\!\big[\tau_{B}D^{(2)}\big],
\qquad
\tau_n^{(2)}
=
\tau_{\mathrm G}\,\tr_S\!\big(\tau^{(2)}\big),
\end{equation}
where $\tau_B={e^{-\beta H_B}}/{\tr_B[e^{-\beta H_B}]}$ is the thermal equilibrium state of the bath and
\begin{equation}
D^{(2)}
=
-\int_{0}^{-i\beta}\! dt'_1 \int_{0}^{t'_1}\! dt'_2\;
\tilde{H}_{SB}(t'_1)\,\tilde{H}_{SB}(t'_2),
\label{eq:D2}
\end{equation}
is the second-order imaginary-time-ordered contribution. The parameters $t'_1$ and $t'_2$ denote two points on the same imaginary-time contour.
The interaction-picture operator on this contour is
\begin{equation}
\tilde{H}_{SB}(t')=e^{\,i(H_S+H_B)t'}\,H_{SB}\,e^{-\,i(H_S+H_B)t'},
\label{eq:HSB_imag_time}
\end{equation}
with $t'\in[0,-i\beta]$.
Collecting terms of order $H_{SB}^{(2)}$ results in the second-order stationarity condition
\begin{equation}
-i[H_S,\Delta\tau_{\mathrm{MF}}^{(2)}] + \mathcal R^{(2)}[\tau_G]=0.
\label{eq:stationarity_order2}
\end{equation}
Since $\tau_n^{(2)}\propto\tau_{\mathrm G}$ commutes with $H_S$, one finds
\begin{equation}
-i[H_S,\tau^{(2)}] + \mathcal R^{(2)}[\tau_{\mathrm G}]=0.
\label{eq:stationarity_order22}
\end{equation}
For off-diagonal elements ($n\neq m$), Eq.~\eqref{eq:stationarity_order22} determines the coherence corrections
\begin{equation}
[{\tau^{(2)}}]_{nm}=\frac{1}{i\delta_{nm}}\big(\mathcal R^{(2)}[\tau_G]\big)_{nm},
\qquad n\neq m.
\label{eq:rho2_coherences_from_R2}
\end{equation}
\blue{
Equation~\eqref{eq:rho2_coherences_from_R2} gives the
second-order equilibrium coherence obtained from the perturbative
stationarity condition, showing that the coherence sector is
determined at order $H_{SB}^{(2)}$. In the asymptotic Markovian
limit, the propagated CCQME steady state satisfies
\begin{equation}
0=
-i\Big[H_S,\big[\rho(t\to\infty)\big]_\text{CCQME}\Big]
+
\mathcal R^{(2)}
\!\left[
\big[\rho(t\to\infty)\big]_\text{CCQME}
-
\mathcal C^{(2)}\!\left[\big[\rho(t\to\infty)\big]_\text{CCQME}\right]
\right].
\label{eq:ccqme_ss_condition_appendix}
\end{equation}
For an off-diagonal element, one arrives at
\begin{equation}
\Big[\big[\rho(t\to\infty)\big]_\text{CCQME}\Big]_{nm}
=
\frac{
\left[
\mathcal R^{(2)}
\!\left[
\big[\rho(t\to\infty)\big]_\text{CCQME}
-
\mathcal C^{(2)}\!\left[\big[\rho(t\to\infty)\big]_\text{CCQME}\right]
\right]
\right]_{nm}
}{
i\delta_{nm}
},
\qquad n\neq m .
\label{eq:ccqme_ss_coherence_appendix}
\end{equation}
This expression is not identical to
Eq.~\eqref{eq:rho2_coherences_from_R2}.
From the unnormalized part of
Eq.~\eqref{eq:rhoMFviaQ2} one obtains
\begin{equation}
\big[\rho(t\to\infty)\big]_\text{CCQME}
\simeq
(1+\mathcal C^{(2)})[\tau_G].
\label{eq:ccqme_ss_pert_appendix}
\end{equation}
Hence
\begin{equation}
(1-\mathcal C^{(2)})\Big[\big[\rho(t\to\infty)\big]_\text{CCQME}\Big]
\simeq
(1-\mathcal C^{(2)})(1+\mathcal C^{(2)})[\tau_G]
=
\tau_G
-
\mathcal C^{(2)}\mathcal C^{(2)}[\tau_G].
\label{eq:inverse_correction_appendix}
\end{equation}
The last term is fourth order in the system–bath coupling and is consequently omitted in a second-order theoretical framework. Following this truncation, Eq.~\eqref{eq:ccqme_ss_coherence_appendix} reduces to the static equilibrium coherence described in Eq.~\eqref{eq:rho2_coherences_from_R2}. As a result, any deviations between the propagated non-secular CCQME coherence and the second-order mean-force coherence should be attributed to finite-coupling higher-order corrections.}

By contrast, the same second-order condition does not determine the second-order population correction $[{\tau^{(2)}}]_{nn}$ ~\cite{Becker2022, BeckerDiss2022}. To be precise, 
In the diagonal sector ($n=m$) the commutator vanishes identically, and Eq.~\eqref{eq:stationarity_order2} reduces to the population constraint
\begin{equation}
0=\big(\mathcal R^{(2)}[\tau_G]\big)_{nn},
\label{eq:population_constraint_order2}
\end{equation}
fixing only the Gibbs state (zeroth order), represented in Eq.~\eqref{eq:rho0}. 
The missing information enters the diagonal stationarity condition only at the next order in the consistent hierarchy. 
At order $H_{SB}^{(4)}$, one obtains
\begin{equation}
0=\big(\mathcal R^{(2)}[\Delta\tau_{\mathrm{MF}}^{(2)}]\big)_{nn}
+\big(\mathcal R^{(4)}[\tau_{\mathrm G}]\big)_{nn}.
\label{eq:population_constraint_order4}
\end{equation}
signifying that 
reproducing equilibrium
populations correctly up to $\mathcal O(H_{SB}^{(2)})$ requires information from the fourth-order contribution
$\mathcal R^{(4)}$ to the time-local generator.

\onecolumngrid
\section{Full Redfield superoperator and CCQME for a two--level system}\label{Redfieldfulll_derivation}
\section*{}

This appendix implements the full non-secularized Redfield formalism and the corresponding CCQME in the same non-secular framework for a two--level subsystem of the TAA molecule with energy eigenstates
$|0\rangle$ and $|1\rangle$. For the two--level system with $H_S|n\rangle = E_n|n\rangle$,
the system Hamiltonian is
\[
H_S \;=\; E_0\,|0\rangle\langle 0| \;+\; E_1\,|1\rangle\langle 1|,
\]
and in the matrix form gives
\[
H_S \;=\;
\begin{pmatrix}
E_0 & 0 \\
0   & E_1
\end{pmatrix}.
\]
The Redfield master equation for the reduced density matrix  is described as  
\begin{equation}
\frac{\partial \rho(t)}{\partial t}
=
-i[H_S,\rho(t)] + \mathcal R^{(2)}[\rho(t)],
\label{eq:full_Redfield_ME}
\end{equation}
where 
\begin{equation}
\rho(t)=
\begin{pmatrix}
\rho_{00}(t) & \rho_{01}(t)\\[1mm]
\rho_{10}(t) & \rho_{11}(t)
\end{pmatrix},
\label{mmatrix}
\end{equation}
is the reduced density matrix of the subsystem and 
$\mathcal R^{(2)}[\rho]$ is the Born--Markov Redfield superoperator.
In the energy basis $\{|0\rangle,|1\rangle\}$, the Hamiltonian is diagonal, so that the unitary evolution reads
\[
[H_{S},\rho(t)]_{00} = [H_{S},\rho(t)]_{11} = 0,
\quad
[H_{S},\rho(t)]_{01} = (E_0-E_1)\rho_{01}(t) = \delta_{01}\,\rho_{01}(t),
\quad
[H_{S},\rho(t)]_{10} = (E_1-E_0)\rho_{10}(t) = \delta_{10}\,\rho_{10}(t),
\]
where $\delta_{10} = E_1-E_0>0$ and $\delta_{01} = -\delta_{10}$ are the Bohr frequencies.
Hence the unitary part acts only on the coherences
\begin{equation}
-i[H_{S},\rho(t)]
=
-i
\begin{pmatrix}
0 & \delta_{01}\,\rho_{01}(t)\\[2pt]
\delta_{10}\,\rho_{10}(t) & 0
\end{pmatrix}.
\label{eq:unitary_block_2level}
\end{equation}
 The Redfield superoperator defined in
Eq.~\eqref{eq:appB_R_def} acts on $\rho(t)$ with matrix elements
\begin{equation}
\label{eq:matrix-action}
\big(\mathcal R^{(2)}[\rho(t)]\big)_{pn}
=
\sum_{k,m}\Big[
{\cal{K}}_{pk}\,\rho_{km}(t)\,q_{mn}
- q_{pk}\,{\cal{K}}_{km}\,\rho_{mn}(t)
+ q_{pk}\,\rho_{km}(t)\,({\cal{K}}^\dagger)_{mn}
- \rho_{pk}(t)\,({\cal{K}}^\dagger)_{km}\,q_{mn}
\Big],
\end{equation}
where $\mathcal{K}$ is the time-independent convolution operator expressed in
Eq.~\eqref{eq:appB_K_def} and $q$ is the proton-transfer coordinate operator given in Eq.~\eqref{eq:q_operator}.
For the two--level truncation
\begin{equation}
q =
\begin{pmatrix}
q_{00} & q_{01} \\
q_{10} & q_{11},
\end{pmatrix},
\end{equation}
where $q$ is real--symmetric so that $ q_{00},q_{11}\in\mathbb{R}$ and  $q_{10}=q_{01}\in\mathbb{R}$
(see Table~\ref{tab:matrixelements}). 
We introduce the following notation for $\mathbb{T}(\delta_{nm})$ expressed in Eq.~\eqref{tunnelingrate_matsubara} as
\begin{equation}
\mathbb{T}_{0} \equiv \mathbb{T}(0), \qquad
\mathbb{T}(+\delta) \equiv \mathbb{T}(\delta_{10}),\qquad
\mathbb{T}(-\delta) \equiv \mathbb{T}(\delta_{01}).
\end{equation}
Using the Markovian convolution operator $\mathcal{K}$ defined in Eq.~\eqref{convoluted_1} and Eq.~\eqref{tunneling_rate_2},
the two--level matrices read
\begin{equation}
\mathcal{K} =
\begin{pmatrix}
q_{00}\,\mathbb{T}_{0} & q_{01}\,\mathbb{T}(-\delta) \\
q_{10}\,\mathbb{T}(+\delta) & q_{11}\,\mathbb{T}_{0}
\end{pmatrix},
\qquad
\mathcal{K}^\dagger =
\begin{pmatrix}
q_{00}\,\mathbb{T}_{0}^\ast & q_{10}\,\mathbb{T}(+\delta)^\ast \\
q_{01}\,\mathbb{T}(-\delta)^\ast & q_{11}\,\mathbb{T}_{0}^\ast
\end{pmatrix}.
\end{equation}
For convenience, we decompose the Redfield superoperator in Eq.~\eqref{eq:matrix-action} as
\[
\mathcal{R}^{(2)}[\rho(t)] = A + B - C - D,
\]
with
\[
A = \mathcal{K}\rho(t)\, q,\qquad
B = q\,\rho(t)\,\mathcal{K}^\dagger,\qquad
C = q\,\mathcal{K}\rho(t),\qquad
D = \rho(t)\,\mathcal{K}^\dagger q.
\]

\subsection*{1.1 The matrix $A=\mathcal{K}\rho q$}

We first evaluate the contribution $A=\mathcal{K}\rho(t)\,q$ 
\[
A =
\begin{pmatrix}
A_{00} & A_{01} \\
A_{10} & A_{11}
\end{pmatrix}.
\]
Using the definition of $\mathcal{K}$ given above, we obtain
\begin{equation}
\begin{aligned}
A_{00} &=
\mathbb T(0)\Big(q_{00}^{2}\rho_{00}(t)+q_{00}\boldsymbol{q}\,\rho_{01}(t)\Big)
+\mathbb T(-\delta)\Big(\boldsymbol{q}q_{00}\rho_{10}(t)+\boldsymbol{q}^{2}\rho_{11}(t)\Big),
\\
A_{01} &=
\mathbb T(0)\Big(q_{00}\boldsymbol{q}\,\rho_{00}(t)+q_{00}q_{11}\rho_{01}(t)\Big)
+\mathbb T(-\delta)\Big(\boldsymbol{q}^{2}\rho_{10}(t)+\boldsymbol{q}q_{11}\rho_{11}(t)\Big),
\\
A_{10} &=
\mathbb T(+\delta)\Big(\boldsymbol{q}q_{00}\rho_{00}(t)+\boldsymbol{q}^{2}\rho_{01}(t)\Big)
+\mathbb T(0)\Big(q_{11}q_{00}\rho_{10}(t)+q_{11}\boldsymbol{q}\,\rho_{11}(t)\Big),
\\
A_{11} &=
\mathbb T(+\delta)\Big(\boldsymbol{q}^{2}\rho_{00}(t)+\boldsymbol{q}q_{11}\rho_{01}(t)\Big)
+\mathbb T(0)\Big(q_{11}\boldsymbol{q}\,\rho_{10}(t)+q_{11}^{2}\rho_{11}(t)\Big),
\end{aligned}
\end{equation}
where $\boldsymbol{q}=q_{01}=q_{10}$ for the real--symmetric two--level proton-transfer coordinate.
\subsection*{1.2 The matrix $B=q\,\rho(t)\,\mathcal{K}^\dagger$}

Next, we evaluate $B=q\,\rho(t)\,\mathcal{K}^\dagger$
\[
B =
\begin{pmatrix}
B_{00} & B_{01} \\
B_{10} & B_{11}
\end{pmatrix},
\]
which yields
\begin{equation}
\begin{aligned}
B_{00} &=
\mathbb{T}(0)^{\ast}\!\Big(q_{00}^{2}\rho_{00}(t)+\boldsymbol{q}\,q_{00}\rho_{10}(t)\Big)
+\mathbb{T}(-\delta)^{\ast}\!\Big(\boldsymbol{q}\,q_{00}\rho_{01}(t)+\boldsymbol{q}^{2}\rho_{11}(t)\Big),
\\
B_{01} &=
\mathbb{T}(+\delta)^{\ast}\!\Big(\boldsymbol{q}\,q_{00}\rho_{00}(t)+\boldsymbol{q}^{2}\rho_{10}(t)\Big)
+\mathbb{T}(0)^{\ast}\!\Big(q_{00}q_{11}\rho_{01}(t)+\boldsymbol{q}\,q_{11}\rho_{11}(t)\Big),
\\
B_{10} &=
\mathbb{T}(0)^{\ast}\!\Big(\boldsymbol{q}\,q_{00}\rho_{00}(t)+q_{00}q_{11}\rho_{10}(t)\Big)
+\mathbb{T}(-\delta)^{\ast}\!\Big(\boldsymbol{q}^{2}\rho_{01}(t)+\boldsymbol{q}\,q_{11}\rho_{11}(t)\Big),
\\
B_{11} &=
\mathbb{T}(+\delta)^{\ast}\!\Big(\boldsymbol{q}^{2}\rho_{00}(t)+\boldsymbol{q}\,q_{11}\rho_{10}(t)\Big)
+\mathbb{T}(0)^{\ast}\!\Big(\boldsymbol{q}\,q_{11}\rho_{01}(t)+q_{11}^{2}\rho_{11}(t)\Big).
\end{aligned}
\end{equation}

\subsection*{1.3 The matrix $C = q\,\mathcal{K}\rho(t)$}

The third contribution is $C=q\,\mathcal{K}\rho(t)$
\[
C =
\begin{pmatrix}
C_{00} & C_{01} \\
C_{10} & C_{11}
\end{pmatrix},
\]
with
\begin{equation}
\begin{aligned}
C_{00}
&=
\mathbb T(0)\Big(q_{00}^{2}\rho_{00}(t)+\boldsymbol{q}\,q_{11}\rho_{10}(t)\Big)
+\mathbb T(+\delta)\Big(\boldsymbol{q}^{2}\rho_{00}(t)\Big)
+\mathbb T(-\delta)\Big(\boldsymbol{q}\,q_{00}\rho_{10}(t)\Big),
\\
C_{01}
&=
\mathbb T(0)\Big(q_{00}^{2}\rho_{01}(t)+\boldsymbol{q}\,q_{11}\rho_{11}(t)\Big)
+\mathbb T(+\delta)\Big(\boldsymbol{q}^{2}\rho_{01}(t)\Big)
+\mathbb T(-\delta)\Big(\boldsymbol{q}\,q_{00}\rho_{11}(t)\Big),
\\
C_{10}
&=
\mathbb T(0)\Big(\boldsymbol{q}\,q_{00}\rho_{00}(t)+q_{11}^{2}\rho_{10}(t)\Big)
+\mathbb T(+\delta)\Big(\boldsymbol{q}\,q_{11}\rho_{00}(t)\Big)
+\mathbb T(-\delta)\Big(\boldsymbol{q}^{2}\rho_{10}(t)\Big),
\\
C_{11}
&=
\mathbb T(0)\Big(\boldsymbol{q}\,q_{00}\rho_{01}(t)+q_{11}^{2}\rho_{11}(t)\Big)
+\mathbb T(+\delta)\Big(\boldsymbol{q}\,q_{11}\rho_{01}(t)\Big)
+\mathbb T(-\delta)\Big(\boldsymbol{q}^{2}\rho_{11}(t)\Big).
\end{aligned}
\end{equation}

\subsection*{1.4 The matrix $D=\rho(t)\,\mathcal{K}^\dagger q$}

Finally, we evaluate $D=\rho(t)\,\mathcal{K}^\dagger q$

\[
D =
\begin{pmatrix}
D_{00} & D_{01} \\
D_{10} & D_{11}
\end{pmatrix},
\]
with the packed expressions
\begin{equation}
\begin{aligned}
D_{00}
&=
\mathbb T(0)^\ast\Big(q_{00}^{2}\rho_{00}(t)+\boldsymbol{q}\,q_{11}\rho_{01}(t)\Big)
+\mathbb T(+\delta)^\ast\Big(\boldsymbol{q}^{2}\rho_{00}(t)\Big)
+\mathbb T(-\delta)^\ast\Big(\boldsymbol{q}\,q_{00}\rho_{01}(t)\Big),
\\
D_{01}
&=
\mathbb T(0)^\ast\Big(\boldsymbol{q}\,q_{00}\rho_{00}(t)+q_{11}^{2}\rho_{01}(t)\Big)
+\mathbb T(+\delta)^\ast\Big(\boldsymbol{q}\,q_{11}\rho_{00}(t)\Big)
+\mathbb T(-\delta)^\ast\Big(\boldsymbol{q}^{2}\rho_{01}(t)\Big),
\\
D_{10}
&=
\mathbb T(0)^\ast\Big(q_{00}^{2}\rho_{10}(t)+\boldsymbol{q}\,q_{11}\rho_{11}(t)\Big)
+\mathbb T(+\delta)^\ast\Big(\boldsymbol{q}^{2}\rho_{10}(t)\Big)
+\mathbb T(-\delta)^\ast\Big(\boldsymbol{q}\,q_{00}\rho_{11}(t)\Big),
\\
D_{11}
&=
\mathbb T(0)^\ast\Big(\boldsymbol{q}\,q_{00}\rho_{10}(t)+q_{11}^{2}\rho_{11}(t)\Big)
+\mathbb T(+\delta)^\ast\Big(\boldsymbol{q}\,q_{11}\rho_{10}(t)\Big)
+\mathbb T(-\delta)^\ast\Big(\boldsymbol{q}^{2}\rho_{11}(t)\Big).
\end{aligned}
\end{equation}

Collecting all terms, the Redfield superoperator in this basis reads
\begin{equation}
\mathcal R^{(2)}[\rho(t)] =
\begin{pmatrix}
R_{00} & R_{01}\\[2pt]
R_{10} & R_{11}
\end{pmatrix}
=
\begin{pmatrix}
A_{00}+B_{00}-C_{00}-D_{00}
&&
A_{01}+B_{01}-C_{01}-D_{01}
\\[4pt]
A_{10}+B_{10}-C_{10}-D_{10}
&&
A_{11}+B_{11}-C_{11}-D_{11}
\end{pmatrix}.
\label{eq:R2_ABCD}
\end{equation}

The full
two--level Redfield master equation in compact form reads
\begin{equation}
\frac{\partial \rho(t)}{\partial t}
=
-i[H_S,\rho(t)]
+
\mathcal R^{(2)}[\rho(t)]
=
-i
\begin{pmatrix}
0 & \delta_{01}\rho_{01}(t)\\[2pt]
\delta_{10}\rho_{10}(t) & 0
\end{pmatrix}
+
\begin{pmatrix}
R_{00} & R_{01}\\[2pt]
R_{10} & R_{11}
\end{pmatrix}.
\end{equation}

\section*{3. Canonical superoperator $\mathcal C^{(2)}$ and CCQME}

We now connect the Redfield superoperator to canonical
superoperator $\mathcal{C}^{(2)}$ and the CCQME.  
The canonical superoperator $\mathcal C^{(2)}$ is defined in Eq.~\eqref{ct}. In the energy eigenbasis $\{|n\rangle\}$, the off--diagonal ($\Pi_{\text{coh}}$) and diagonal ($\Pi_{\text{pop}}$) projectors are defined as
\begin{equation}
\Pi_{\text{coh}}[.] \equiv \sum_{n\neq m} |n\rangle\langle n|\,.\,|m\rangle\langle m|,
\qquad
\Pi_{\text{pop}}[.] \equiv \sum_{n} |n\rangle\langle n|\,.\,|n\rangle\langle n|.
\label{eq:Pi_def_2level}
\end{equation}

\subsection*{3.1 Coherence part of $\mathcal C^{(2)}$ (off-diagonal elements)}
 The coherence (off-diagonal) part of the canonical map is given by the
inverse commutator action on $n\neq m$ matrix elements~\cite{Becker2022, BeckerDiss2022}

\begin{equation}
\big[\mathcal{C}^{(2)}[\rho(t)]\big]_{nm}
=
\frac{1}{i\delta_{nm}}\,
\big[\mathcal{R}^{(2)}[\rho(t)]\big]_{nm},
\qquad n\neq m,
\quad \delta_{nm}=E_n-E_m .
\end{equation}
For N=2 with energy eigenstates
$|0\rangle$ and $|1\rangle$, one obtains
\begin{equation}
\begin{aligned}
\big[\mathcal C^{(2)}[\rho(t)]\big]_{01}
&=
\frac{1}{i\delta_{01}}\big[\mathcal R^{(2)}[\rho(t)]\big]_{01}
=
-\frac{1}{i\delta}\big[\mathcal R^{(2)}[\rho(t)]\big]_{01},\\[4pt]
\big[\mathcal C^{(2)}[\rho(t)]\big]_{10}
&=
\frac{1}{i\delta_{10}}\big[\mathcal R^{(2)}[\rho(t)]\big]_{10}
=
+\frac{1}{i\delta}\big[\mathcal R^{(2)}[\rho(t)]\big]_{10}.
\end{aligned}
\label{eq:C2_coh_2level}
\end{equation}

\subsection*{3.2 Population part of $\mathcal C^{(2)}$ (diagonal elements)}
The population segment of Eq.~\eqref{ct} is described as ~\cite{Becker2022, BeckerDiss2022}
\begin{equation}
\Pi_{\text{pop}}\,\mathcal{C}^{(2)}[\rho]
=
\Pi_{\text{pop}} \sum_{n\neq l} |q_{nl}|^2
\left[
\underbrace{
\frac{\partial}{\partial \delta_{nl}}\,
\Im\!\big[\mathbb{T}(\delta_{nl})\big]\;
\mathcal{D}\!(L)[\rho]
}_{\text{(I) GKSL-type term}}
+
\underbrace{
\Im\!\big[\mathbb{T}(\delta_{ln})\big]\;
\ket{n}\big(\partial_{E_n}\rho_{nn}\big)\bra{n}
}_{\text{(II) derivative term}}
\right].
\label{17}
\end{equation}
Here $\mathcal D(L)$ denotes the GKSL superoperator
\begin{equation}
\mathcal D(L)[\rho]
=
L\rho L^\dagger-\frac12\{L^\dagger L,\rho\},\qquad L=|n\rangle\langle l|,
\label{eq:DKSL_def}
\end{equation}
where $\{A,B\}\equiv AB+BA$ is the anticommutator.
For $N=2$ the only jump operators are
\[
L_{01}=|0\rangle\langle1|,
\qquad
L_{10}=|1\rangle\langle0|,
\qquad
|q_{01}|^2=|q_{10}|^2=\boldsymbol q^{\,2}.
\]
Using Eq.~\eqref{eq:DKSL_def},  one finds
\begin{equation}
\mathcal D(L_{01})[\rho]=
\begin{pmatrix}
\rho_{11} & -\tfrac12\rho_{01}\\[2pt]
-\tfrac12\rho_{10} & -\rho_{11}
\end{pmatrix},
\qquad
\mathcal D(L_{10})[\rho]=
\begin{pmatrix}
-\rho_{00} & -\tfrac12\rho_{01}\\[2pt]
-\tfrac12\rho_{10} & \rho_{00}
\end{pmatrix}.
\end{equation}
Therefore, the population projector $\Pi_{\text{pop}}$ extracts the diagonal
parts
\begin{equation}
\Pi_{\text{pop}}\!\left[\mathcal D(|0\rangle\langle1|)[\rho]\right]
=
\begin{pmatrix}
\rho_{11} & 0\\
0 & -\rho_{11}
\end{pmatrix},
\quad
\Pi_{\text{pop}}\!\left[\mathcal D(|1\rangle\langle0|)[\rho]\right]
=
\begin{pmatrix}
-\rho_{00} & 0\\
0 & \rho_{00}
\end{pmatrix}.
\end{equation}
From Eq.~\eqref{17}, the GKSL-type contribution to the diagonal sector of $\mathcal C^{(2)}$ is
\begin{equation}
\big[\mathcal C^{(2)}[\rho]\big]_{\mathrm{diag}}^{\mathrm{GKSL}}
=
\Pi_{\text{pop}}\sum_{n\neq l}|q_{nl}|^{2}\,
\frac{\partial}{\partial \delta_{nl}}\Im\!\big[\mathbb{T}(\delta_{nl})\big]\;
\mathcal D(L)[\rho].
\label{eq:C2_diag_GKSL_general}
\end{equation}
For $N=2$, only $(n,l)=(0,1)$ and $(1,0)$ contribute. Therefore, for the $00$ component, one obtains
\begin{align}
\big[\mathcal C^{(2)}[\rho]\big]_{00}^{\mathrm{GKSL}}
&=
\boldsymbol q^{\,2}\Bigg[
\frac{\partial}{\partial \delta_{01}}\Im\!\big[\mathbb{T}(\delta_{01})\big]\,
\underbrace{\big(\Pi_{\text{pop}}\mathcal D(L_{01})[\rho]\big)_{00}}_{=\ \rho_{11}}
+
\frac{\partial}{\partial \delta_{10}}\Im\!\big[\mathbb{T}(\delta_{10})\big]\,
\underbrace{\big(\Pi_{\text{pop}}\mathcal D(L_{10})[\rho]\big)_{00}}_{=\ -\rho_{00}}
\Bigg]
\nonumber\\[2pt]
&=
\boldsymbol q^{\,2}\Bigg[
\frac{\partial}{\partial \delta_{01}}\Im\!\big[\mathbb{T}(\delta_{01})\big]\;\rho_{11}
-
\frac{\partial}{\partial \delta_{10}}\Im\!\big[\mathbb{T}(\delta_{10})\big]\;\rho_{00}
\Bigg].
\label{eq:C2_GKSL_00_2level}
\end{align}
Analogously, for the $11$ component
\begin{align}
\big[\mathcal C^{(2)}[\rho]\big]_{11}^{\mathrm{GKSL}}
&=
\boldsymbol q^{\,2}\Bigg[
\frac{\partial}{\partial \delta_{01}}\Im\!\big[\mathbb{T}(\delta_{01})\big]\,
\underbrace{\big(\Pi_{\text{pop}}\mathcal D(L_{01})[\rho]\big)_{11}}_{=\ -\rho_{11}}
+
\frac{\partial}{\partial \delta_{10}}\Im\!\big[\mathbb{T}(\delta_{10})\big]\,
\underbrace{\big(\Pi_{\text{pop}}\mathcal D(L_{10})[\rho]\big)_{11}}_{=\ \rho_{00}}
\Bigg]
\nonumber\\[2pt]
&=
\boldsymbol q^{\,2}\Bigg[
\frac{\partial}{\partial \delta_{10}}\Im\!\big[\mathbb{T}(\delta_{10})\big]\;\rho_{00}
-
\frac{\partial}{\partial \delta_{01}}\Im\!\big[\mathbb{T}(\delta_{01})\big]\;\rho_{11}
\Bigg].
\label{eq:C2_GKSL_11_2level}
\end{align}

The remaining diagonal contribution in Eq.~\eqref{17} is the {derivative term (II)}.  The diagonal part of $\mathcal C^{(2)}$ contains operators of the form
$\ket{n}\big(\partial_{E_n}\rho_{nn}\big)\bra{n}$, where $\partial_{E_n}$ is a derivative superoperator
acting on the population vector $\{\rho_{nn}\}$

\begin{equation}
\partial_{E_n}\rho^{(0)}_{nn}
=
\frac{
\displaystyle\sum_{l\neq n}|q_{nl}|^{2}
\left[
\left(\frac{\partial}{\partial \delta_{nl}}\Re\!\big[\mathbb{T}(\delta_{nl})\big]\right)\rho^{(0)}_{ll}
+
\left(\frac{\partial}{\partial \delta_{ln}}\Re\!\big[\mathbb{T}(\delta_{ln})\big]\right)\rho^{(0)}_{nn}
\right]
}{
\displaystyle\sum_{l\neq n}|q_{ln}|^{2}\,\Re\!\big[\mathbb{T}(\delta_{ln})\big]
}\,,
\label{eq:derivative_superoperator_general}
\end{equation}
where 

\begin{equation}
\rho^{(0)}_{nn}
=
\frac{
\displaystyle\sum_{l\neq n}|q_{nl}|^{2}\,\Re\!\big[\mathbb{T}(\delta_{nl})\big]\;\rho^{(0)}_{ll}
}{
\displaystyle\sum_{l\neq n}|q_{ln}|^{2}\,\Re\!\big[\mathbb{T}(\delta_{ln})\big]
}\,,
\label{eq:stationary_balance_general}
\end{equation}
is the zeroth-order steady-state populations.

For $N=2$, we can write the derivative superoperators in the following form
\begin{equation}
\partial_{E_{0}}\rho_{00}
=
\dfrac{
\dfrac{\partial}{\partial\delta_{01}}\Re[{\mathbb{T}}(\delta_{01})]\,\rho_{11}
+ \dfrac{\partial}{\partial\delta_{10}}\Re[{\mathbb{T}}(\delta_{10})]\,\rho_{00}
}{
\Re[{\mathbb{T}}(\delta_{10})]
},
\qquad
\partial_{E_{1}}\rho_{11}
=
\dfrac{
\dfrac{\partial}{\partial\delta_{10}}\Re[{\mathbb{T}}(\delta_{10})]\,\rho_{00}
+ \dfrac{\partial}{\partial\delta_{01}}\Re[{\mathbb{T}}(\delta_{01})]\,\rho_{11}
}{
\Re[{\mathbb{T}}(\delta_{01})]
}.
\label{eq:derivative-terms}
\end{equation}
Collecting both diagonal contributions, we define
\begin{equation}
\begin{aligned}
Q_{00}
&\equiv
\big[\mathcal C^{(2)}[\rho]\big]_{00}
=
|\boldsymbol{q}|^2\Big[
\frac{\partial}{\partial\delta_{01}}\Im[{\mathbb{T}}(\delta_{01})]\,\rho_{11}
-
\frac{\partial}{\partial\delta_{10}}\Im[{\mathbb{T}}(\delta_{10})]\,\rho_{00}
+
\Im[{\mathbb{T}}(\delta_{10})]\,\partial_{E_{0}}\rho_{00}
\Big],\\[6pt]
Q_{11}
&\equiv
\big[\mathcal C^{(2)}[\rho]\big]_{11}
=
|\boldsymbol{q}|^2\Big[
\frac{\partial}{\partial\delta_{10}}\Im[{\mathbb{T}}(\delta_{10})]\,\rho_{00}
-
\frac{\partial}{\partial\delta_{01}}\Im[{\mathbb{T}}(\delta_{01})]\,\rho_{11}
+
\Im[{\mathbb{T}}(\delta_{01})]\,\partial_{E_{1}}\rho_{11}
\Big].
\end{aligned}
\label{eq:Q_def}
\end{equation}

\subsection*{3.3 CCQME as a full master equation}
In the Markovian limit, CCQME reads
\begin{equation}
\frac{\partial}{\partial t}\big[\rho(t)\big]_\text{CCQME}
=
-i[H_S,\rho(t)]
+
\mathcal R^{(2)}\!\Big[(1-\mathcal C^{(2)})[\rho(t)]\Big].
\label{eq:CCQME_full}
\end{equation}
In analogy with Eq.~\eqref{eq:full_Redfield_ME},  Eq.~\eqref{eq:CCQME_full} consists of the same unitary contribution
$-i[H_S,\rho(t)]$ (\ Eq.~\eqref{eq:unitary_block_2level}) and a canonically corrected dissipative contribution
\begin{equation}
\mathcal R^{(2)}\!\Big[(1-\mathcal C^{(2)})[\rho(t)]\Big].
\label{eq:CCQME_diss_part}
\end{equation}
We define the canonically corrected density matrix as
\begin{equation}
\rho'(t)=(1-\mathcal C^{(2)})[\rho(t)],
\label{eq:rho_prime_def}
\end{equation}
where its diagonal entries are
\[
\rho'_{00}(t)=\rho_{00}(t)-Q_{00},
\qquad
\rho'_{11}(t)=\rho_{11}(t)-Q_{11},
\]
and the off-diagonals follow from Eq.~\eqref{eq:C2_coh_2level} as
\[
\rho'_{01}(t)=\rho_{01}(t)-\big[\mathcal C^{(2)}[\rho(t)]\big]_{01}
=\rho_{01}(t)+\frac{1}{i\delta}\big[\mathcal R^{(2)}[\rho(t)]\big]_{01},
\]
\[
\rho'_{10}(t)=\rho_{10}(t)-\big[\mathcal C^{(2)}[\rho(t)]\big]_{10}
=\rho_{10}(t)-\frac{1}{i\delta}\big[\mathcal R^{(2)}[\rho(t)]\big]_{10}.
\]

In the energy basis, Eq.~\eqref{eq:CCQME_full} can be written directly in matrix form as
\begin{equation}
\frac{\partial}{\partial t}
\begin{pmatrix}
\rho_{00}(t) & \rho_{01}(t)\\[2pt]
\rho_{10}(t) & \rho_{11}(t)
\end{pmatrix}_\text{CCQME}
=
\begin{pmatrix}
\big[\mathcal R^{(2)}[\rho'(t)]\big]_{00}
&
+i\delta\,\rho_{01}(t)+\big[\mathcal R^{(2)}[\rho'(t)]\big]_{01}
\\[6pt]
-i\delta\,\rho_{10}(t)+\big[\mathcal R^{(2)}[\rho'(t)]\big]_{10}
&
\big[\mathcal R^{(2)}[\rho'(t)]\big]_{11}
\end{pmatrix}.
\label{eq:CCQME_matrix_form}
\end{equation}

\section{Numerical details}
\label{NumericalDetails}
\section*{}
We found the eigenstates of the effective system Hamiltonian, using the discrete variable representation (DVR) \cite{Colbert1992}. In particular, we use the scheme for one Cartesian dimension, generally restricted to $q \in (-\infty,\infty)$. The grid spacing $\Delta q$ is the only parameter involved and, in this work chosen according to $121$ grid points with $q \in [-1.5,2.1] \ a_0$\blue{, consistent with previous work in Ref.~\citenum{Fischer2023}.

\newrev{As a practical numerical check for the Gaussian-wavepacket calculation at $\gamma=0.5$, we repeated the calculation with an enlarged system representation, increasing the basis from $N=12$ to $N=24$ and the DVR grid from 121 to 512 points. The resulting populations and coordinate expectation values showed only minor differences. Since the basis size and DVR grid density were increased at the same time, this comparison should be understood as a stability check of the reported benchmark rather than as two separate convergence analyses with respect to N and the DVR discretization. The finite coordinate interval, $(q \in [-1.5,2.1],a_0)$, 
 was kept fixed and was not varied independently.} Moreover, for a given coupling strength $\gamma$ or initialization (Gaussian, $\ket{0}$ or $\ket{1}$ initial states), all three presented methods (CCQME, Redfield and HEOM) use exactly the same system eigenstate basis, which ensures comparability and consistent benchmarks -- the main goal of this work.}

Regarding the HEOM reference calculations, we made use of the QuTip package \cite{qutip5}. In this package, two parameters determine the numerical convergence of the HEOM method, namely the hierarchy depth and the number of bath expansion terms. Fig.~\ref{fig:ap_HEOM_convergence} shows the population dynamics of the very first eigenstate for $\gamma=1.0$. For all simulations in the main text, a depth of $5$ and $2$ expansion terms were used. However, the expansion was combined with the Matsubara terminator, which treats higher-order terms (with comparatively weaker coupling) as additional Lindblad terms \cite{qutip5}. The hierarchy depth of $5$ is well converged beyond the third digit. 


\begin{figure*}[h!]
    \centering
    \includegraphics[width=.9\linewidth]{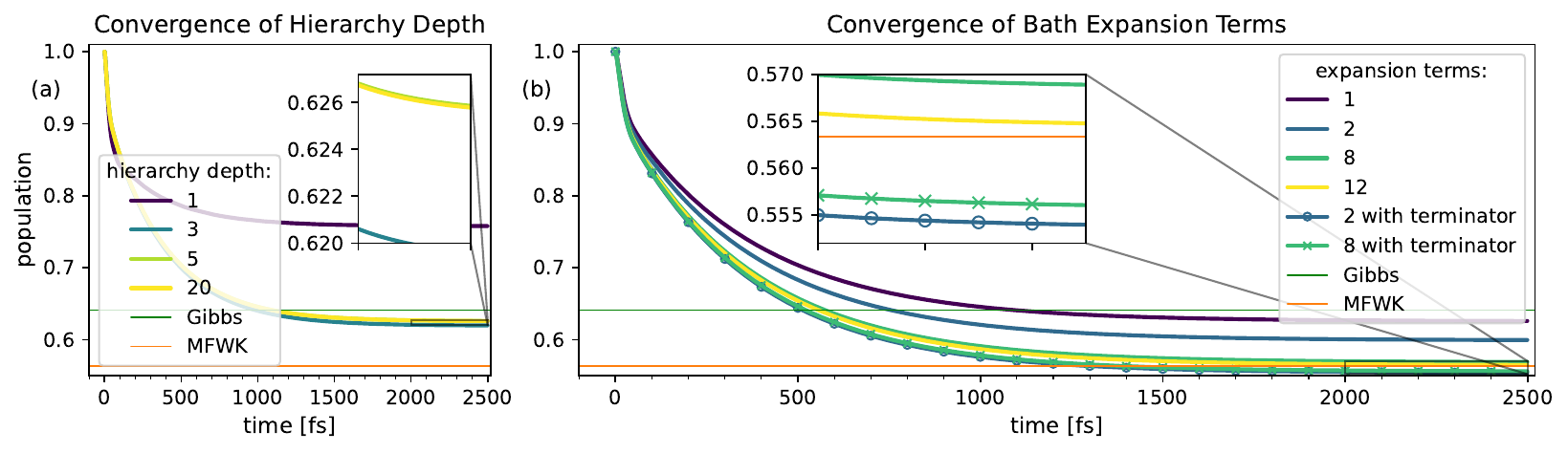}
    \caption{Convergence of HEOM for increasing (a) hierarchy depth and (b) number of bath expansion terms, in case of the ground-state population dynamics for $\gamma=1.0$.}
    \label{fig:ap_HEOM_convergence}
\end{figure*}

\section{Additional population dynamics}
\label{MorePopDynamics}
\section*{}

\begin{figure*}[h!]
\vspace*{-1cm}
    \centering
    \includegraphics[width=.9\linewidth]{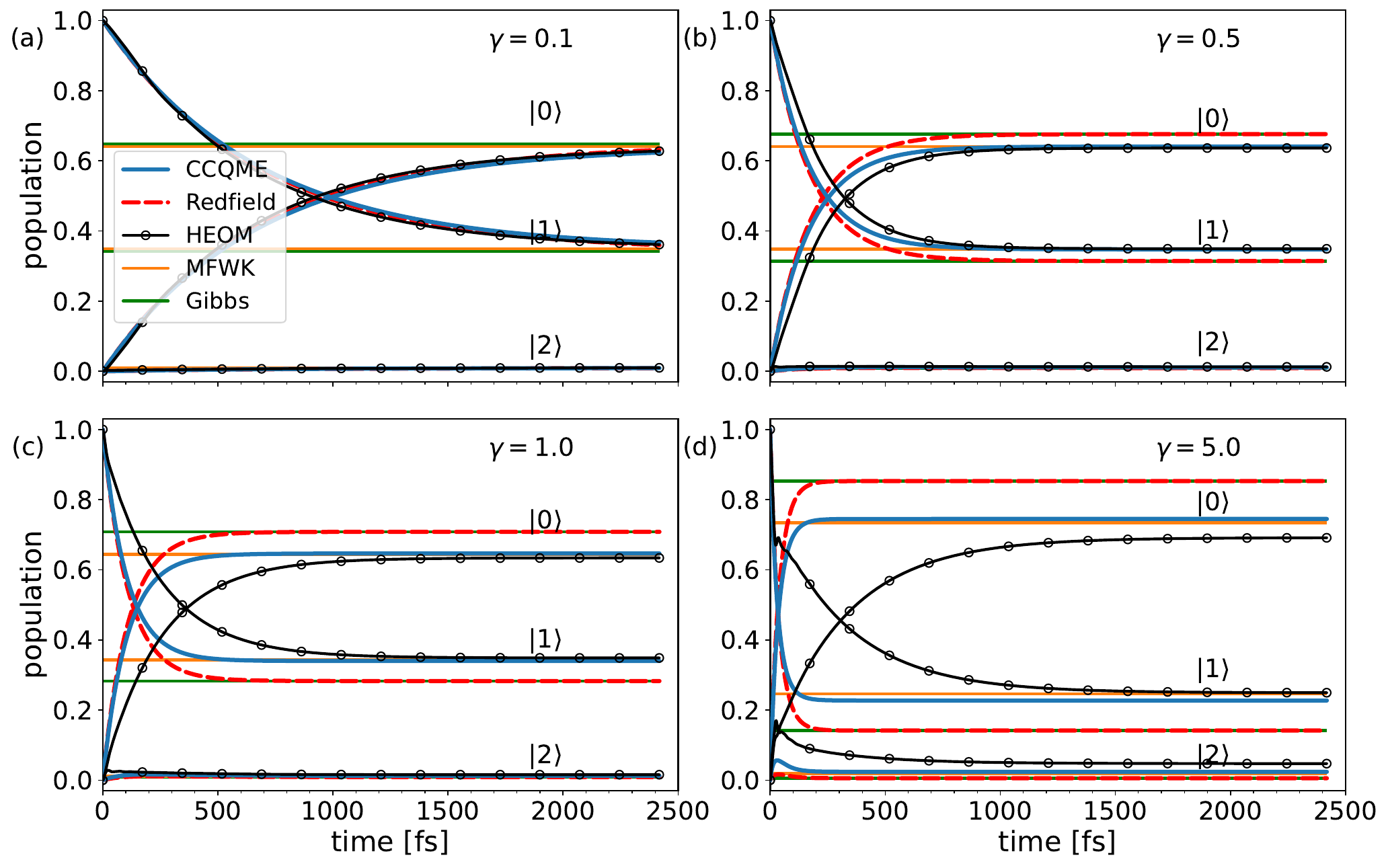}
    \caption{Population dynamics of the three lowest eigenstates of the TAA proton-transfer coordinate, now with the first excited state as initial state and everything else unchanged from Fig.~\ref{FIG1} in the main text.}
    \label{FIG12}
\end{figure*}

Related to Fig.~\ref{FIG1} in the main text, we show in Fig.~\ref{FIG12} additional population dynamics for the same model and parameters; however, with the first excited state $\ket{1}$ as initial state, to observe vibrational relaxation. While the deviations of the mean-force Gibbs state from the regular one remain unchanged, the Redfield equation clearly overestimates relaxation rates, especially for $\gamma=1.0$. On the other hand, CCQME still delivers reasonable results compared to HEOM. This deviation in rates was less significant for $\ket{0}$ as initial state.

\section{Additional wavepacket dynamics}
\label{MoreWPDynamics}
\section*{}

\blue{In the following three figures, we} show additional wavepacket dynamics for increasing coupling strength $\gamma$ to illustrate the influence of the coupling to the bath on the initially coherent oscillations observed in the expected position dynamics. 

\begin{figure}[hbt!]
    \centering
    \includegraphics[width=0.86\linewidth]{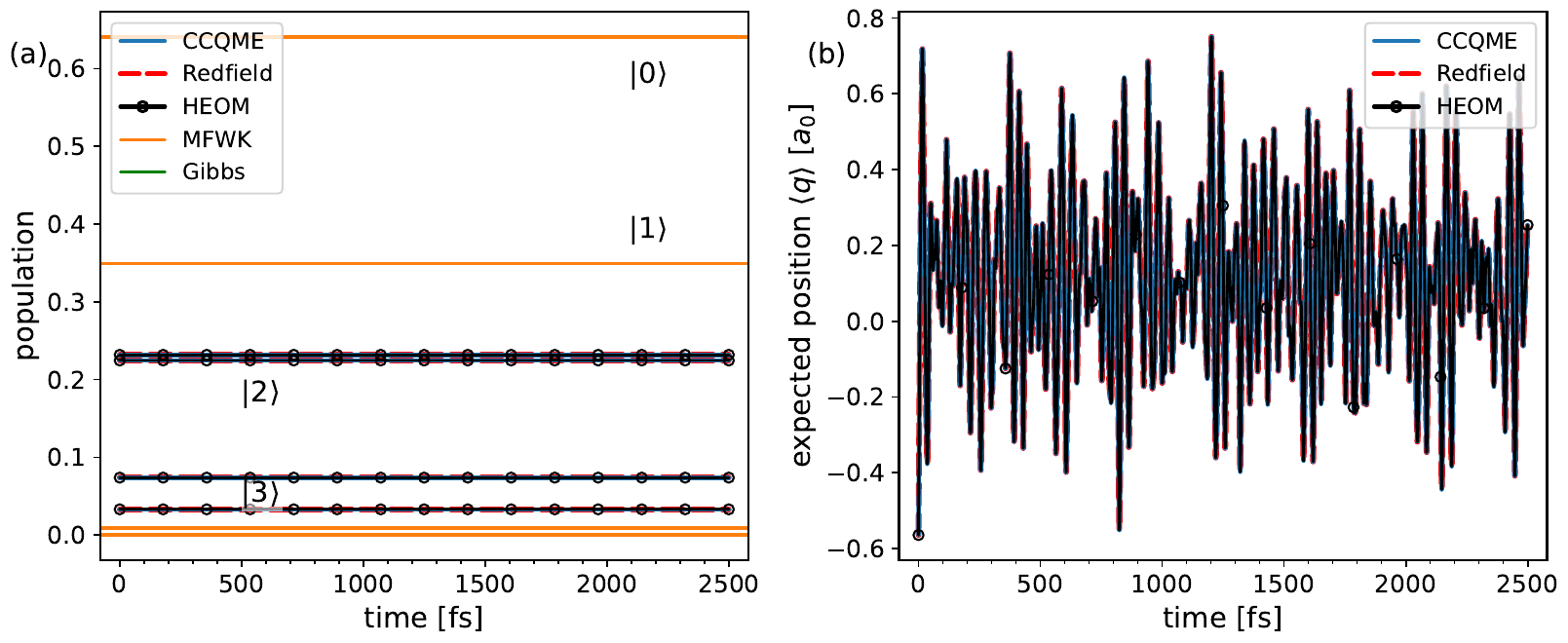}
    \caption{Population dynamics of the four lowest eigenstates and corresponding coordinate expectation values $\expval{q}(t)$ for a Gaussian wavepacket for coupling strength $\gamma=0.0$. Remaining parameters are equal to the main text.}
    \label{fig:wp_dyn_g00}
\end{figure}
\begin{figure}[h!]
    \centering
    \includegraphics[width=0.86\linewidth]{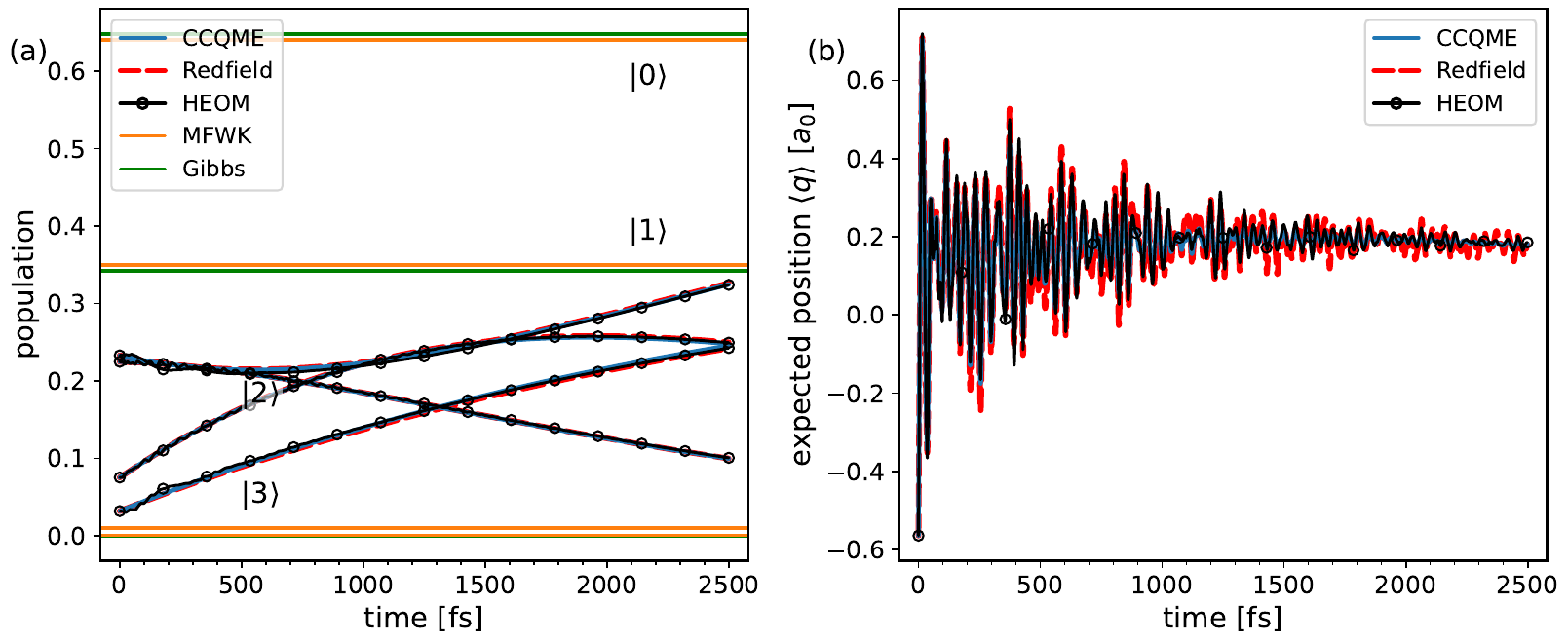}
    \caption{Population dynamics of the four lowest eigenstates and corresponding coordinate expectation values $\expval{q}(t)$ for a Gaussian wavepacket for coupling strength $\gamma=0.1$.}
    \label{fig:wp_dyn_g01}
\end{figure}
\begin{figure}[h!]
    \centering
    \includegraphics[width=0.86\linewidth]{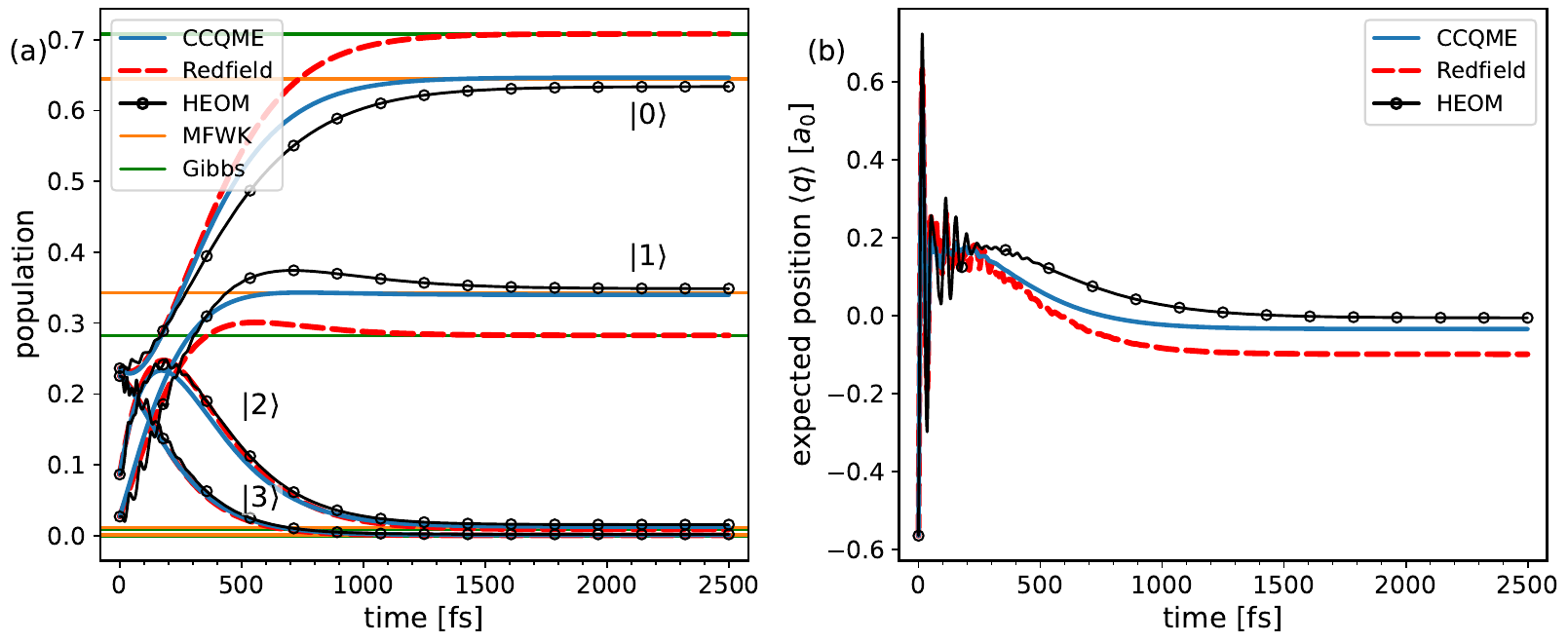}
    \caption{Population dynamics of the four lowest eigenstates and corresponding coordinate expectation values $\expval{q}(t)$ for a Gaussian wavepacket for coupling strength $\gamma=1.0$.}
    \label{fig:wp_dyn_g10}
\end{figure}

\clearpage
\section{Additional Benchmarks}
\label{WPDynBench}
\section*{}

\blue{In Fig.~\ref{fig:wp_dyn_bench}, we provide errors of the transfer coordinate for a Gaussian wavepacket as initial state by comparing the three employed methods (CCQME, Redfield, HEOM) pairwise.}

\begin{figure}[h!]
    \centering
    \includegraphics[width=0.5\linewidth]{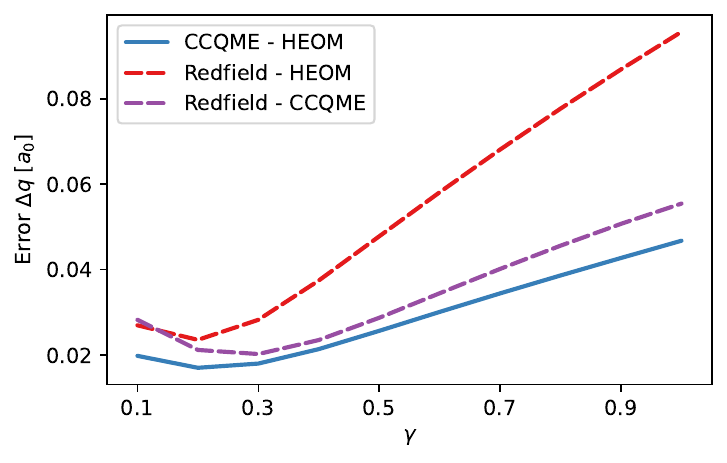}
    \caption{\blue{Time-averaged transfer coordinate error $\Delta q$ (in $a_0$) as a function of the system--bath coupling strength $\gamma$ for a Gaussian wavepacket as initial state, comparing the CCQME and Redfield predictions with the numerically exact HEOM benchmark. The blue solid curve shows $\Delta$ (CCQME--HEOM), the red dashed curve shows $\Delta$ (Redfield--HEOM), and the purple dash--dotted curve shows $\Delta$ (Redfield--CCQME). The errors are averaged over the propagation time window $0$--$2.5$ ps.}}
    \label{fig:wp_dyn_bench}
\end{figure}

\bibliography{Paper_Draft/references}
\bibliographystyle{apsrev4-1}
\end{document}